# The Portevin-Le Chatelier Effect and Beyond

## Mikhail A. Lebyodkin[1,*] and Tatiana A. Lebedkina[1,2,3]

[1] Laboratoire d'Etude des Microstructures et de Mécanique des Matériaux (LEM3), CNRS, Université de Lorraine, Arts & Métiers ParisTech, 7 rue Félix Savart, 57070 Metz, France

[2] Center of Excellence "LabEx DAMAS", Université de Lorraine, 7 rue Félix Savart, 57070 Metz, France

[3] IRT M2P, 4 rue Augustin Fresnel, 57070 Metz, France

**\*** Correspondence: mikhail.lebedkin@univ-lorraine.fr
Tel.: +33 (0)3-72-74-77-71

*This chapter, which discusses the phenomenon of unstable plastic deformation in traditional, basically binary alloys, was prepared for a book devoted to multicomponent high-entropy materials. As such modern materials are also prone to plastic instability, the objective of the chapter was to provide links between these two fields of investigation and allow for a systemic view of the studied phenomenon. The following "pre-preface" marked in italic is aimed at putting the chapter into a proper context.*

## *Pre-Preface*

*Although the notion of "plastic flow" evokes a smooth process, agreeing with one's everyday experience and, more specifically, with the common examples of smooth deformation curves shown to students during the lessons on mechanics, plastic deformation often proceeds in an intermittent manner. The serrated flow is one of the striking features of plastic flow in solids, which reveals a self-organized nature of the dynamics of crystal defects and unifies the problems of plasticity with diverse phenomena observed in complex systems of various nature.*

*While such an instability of smooth plastic flow may be caused by different mechanisms and can occur in various materials, the most abundant examples of this phenomenon have been documented for a wide range of alloys, for which the discovery of intermittent deformation will soon celebrate the 200-year anniversary. It is therefore not surprising that the serrated flow is also an essential feature of the*

*deformation behavior exhibited by high-entropy alloys and has quickly attracted the attention of the researcher community.*

*To provide a comprehensive approach to this problem, the further manuscript will present the serrated flow caused by the Portevin-Le Chatelier effect in conventional low-entropy alloys. It is caused by the interaction of dislocations with solute atoms, which is dynamic in the sense that the solutes do not simply represent immobile obstacles but diffuse and form clouds on the dislocations, so that the respective pinning force depends on the dynamics of all actors. The attention to this effect in the present book is due to numerous proofs that although the subdivision into basic and solute elements is not evident for high-entropy materials, the Portevin-Le Chatelier instability seems to be the mechanism of serrated flow in such alloys in a wide range of experimental conditions. Moreover, this effect served as a model object for the elaboration of various mathematical approaches to testing the complexity of distinct behaviors of plastic deformation. Some of them will be presented in detail in the chapter. As the respective literature is huge, it will not be reviewed systematically. Instead, the authors put an accent on providing the reader with a qualitative knowledge of the basic dynamical regimes uncovered by virtue of the analysis of experimental data obtained on multiple scales.*

## Introduction

This chapter presents a review of investigations into the complexity of plastic flow associated with serrated deformation, or jerky flow, in traditional alloys that have basic elements determining the crystal lattice of the material. The phenomenon of serrated deformation has been known for more than a century [1-3]. This type of dynamical behavior was early understood as resulting from the collective motions of very large groups of crystal defects, notably dislocations [4]. However, approaching the dynamical mechanisms of such processes has only become possible after the occurrence of the theory of nonlinear dissipative systems [5,6] and understanding that collective deformation processes are analogous to self-organization phenomena in dynamical systems abundant both in nature (physics, biology, chemistry…) and in human society (sociology, market, road traffic…) [7-10]. From the viewpoint of the theory of plasticity, the problem is yet larger than that of the macroscopically-serrated deformation. Self-organization appears to be a generic property of plastic deformation on a mesoscopic scale even when deformation curves of bulk samples are smooth, be it the case of pure crystals or, in certain conditions, alloys. These collective phenomena were revealed by virtue of higher-resolution methods, such as acoustic emission or local extensometry [11-13]. Mesoscopic effects also inevitably show up when the sample size is extremely reduced, e.g., in tensile tests on thin wires or compression tests on micropillars [14,15]. Vice versa, abrupt macroscopic stress/strain fluctuations may occur not only in alloys but also in pure materials, and may be caused by different mechanisms, e.g., twinning or catastrophic slip at low temperature [16-21]. Investigations of these phenomena in the spirit of nonlinear dynamical systems commenced in the 1980s and allowed for testing various approaches to the analysis of jerky flow. This experience will obviously be useful for the progress in the understanding of similar phenomena in modern materials with complex microstructures. The field of research is vast and cannot be presented in a single chapter. The following review will cover some aspects of experimental investigation of both macro- and mesoscopic scale effects in the conditions of jerky flow in alloys, well-known as the Portevin-Le Chatelier (PLC) effect [3,4,7,8] and occupying a particular place in the study of self-organization phenomena in plasticity. Finally, a short insight into the mesoscopic-scale complexity will be provided in view of the future directions of research.

## Table of Contents

# 1. Portevin-Le Chatelier effect

The jerky flow manifests itself by recurring localizations of plastic strain in deformation bands, so intense that even the total plastic strain rate, $\dot{\varepsilon}$, referring to the entire specimen significantly exceeds the applied strain rate, $\dot{\varepsilon}_a$, during the band's lifetime. This plastic instability may occur in both interstitial and substitutional alloys including numerous industrial materials, such as steels or Al-based alloys. Its impact on practical application of alloys cannot be overestimated. It may lead to a reduced ductility and irremediable roughness of the rolled sheets, affect the work-hardening behavior or change the fracture type from ductile to brittle; Inversely, it may improve the material strength [4,22,23].

As there exists some discrepancy in the historical aspects, it is important to clarify that although various kinds of such instability are often called after A. Portevin et F. Le Chatelier, who published the first paper on jerky flow in constant strain-rate conditions [3], the very first observations of plastic instability were reported by F. Savart et M.A. Masson for creep tests at loads increased in incremental steps [1,2]. Accordingly, the jerky flow occurring in constant stress-rate tests is often referred to as the Savart-Masson effect. It is the PLC instability that has become a model object for the study of collective effects in plastic deformation. One of advantages of the PLC effect is to provide large amounts of data because the constant-$\dot{\varepsilon}_a$ loading mode allows for unloading due to the elastic reaction of the deformation machine to an abrupt change in the strain of the sample, thus leading to the cessation of the deformation band development. Such stress relaxation makes the instability less crucial and allows for the accumulation of hundreds and even thousands of instability events before fracture or, more precisely, before the onset of necking in the sample. Such a scenario thus provides a basis for analysis by various statistical methods allowing to assess signatures of collective dynamics. In contrast, the sustained stress rate leads to huge strain bursts resulting in the sample failure after several instability events. Typical serrated stress-strain dependences conditioned by the PLC effect are illustrated in Figure 1 for an Al3%Mg alloy [24]. Figure 2 presents an example of a photograph of traces of deformation bands on the surface of a tensile specimen of an Al5%Mg alloy [25].

Moreover, as can be readily recognized in Figure 1, the manifestations of the PLC effect drastically depend on the deformation conditions, e.g., on $\dot{\varepsilon}_a$. As will be illustrated in this Chapter, the PLC effect brings one of the richest examples of complex behavior associated with transitions between distinct dynamical regimes characteristic of nonlinear systems of various nature, such as deterministic chaos [26], self-organized criticality (SOC) [27], or synchronization [28]. This diversity gives a general and multidisciplinary scope to the problem of plastic instability, evoking the problems arising from various fields of research.

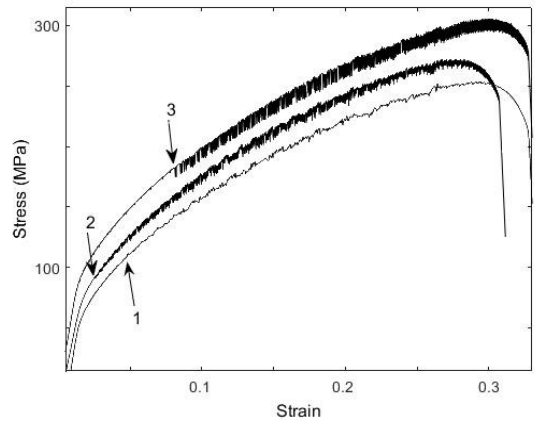

*Figure 1. Portions of tensile curves of an AlMg alloy (room temperature), presenting three types of stress serrations commonly distinguished for the PLC effect: (1) type A, $\dot{\varepsilon}_a = 2 \times 10^{-3}\,s^{-1}$; (2) type B, $\dot{\varepsilon}_a = 2 \times 10^{-4}\,s^{-1}$; (3) type C, $\dot{\varepsilon}_a = 6.7 \times 10^{-5}\,s^{-1}$. Arrows indicate the critical strain $\varepsilon_{cr}$ for the onset of instability.*

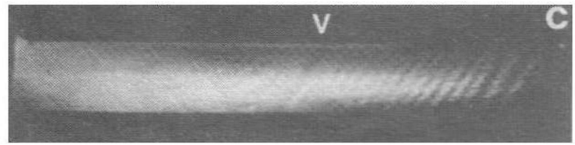

*Figure 2. Traces of PLC bands observed on the surface of an AlMg alloy deformed in the conditions of type B behavior at $\dot{\varepsilon}_a = 10^{-3}\,s^{-1}$. The specimen width is 5 mm. The PLC bands can be seen with a naked eye on the specimen surface. Their typical width reported in the literature varies from several hundred μm to about ten mm (Figure from Ref. [25]).*

More specifically, the PLC effect has important advantages for the study of self-organization in dislocation ensembles. First of all, since the relevant collective effects manifest themselves on a macroscopic scale, the combination of traditional mechanical tests with higher-resolution techniques allows for getting access to collective processes in a wide range of resolution. Furthermore, the PLC effect only appears after a certain critical strain, $\varepsilon_{cr}$, as indicated by arrows in Figure 1. This delay makes it possible to compare fine-scale behaviors during a "homogeneous deformation" stage, then during the instability. Finally, apart from its own interest, the PLC effect serves as a model object for comparison with fine-scale plasticity in materials non-subject to macroscopic instability.

The microscopic mechanism of plastic instability is generally attributed to the so-called dynamical strain ageing (DSA), i.e., additional pinning of mobile dislocations due to solute atoms diffusing in the dislocation elastic field [29-32]. It is noteworthy that non-diffusional models have also been suggested and may present an alternative explanation in a range of sufficiently high $\dot{\varepsilon}_a$ values [33]. Without a loss of generality, the interpretation of stress drops is based on the concept of negative strain-rate sensitivity of stress (SRS) in a certain strain-rate interval (cf. Figure 1), giving rise to an $N$-shaped $\sigma(\dot{\varepsilon})$ dependence [34-38]. The left chart in Figure 3 illustrates the occurrence of this dependence. The early interpretation of dynamical behavior in this framework had a local character in the sense that the same stress value was considered to act over the entire gage length of the sample, i.e., the spatial heterogeneity was disregarded. Under this approximation, the nonlinear SRS leads to the so-called relaxation oscillations, a well-known type of instability, notably, in electronics [39].

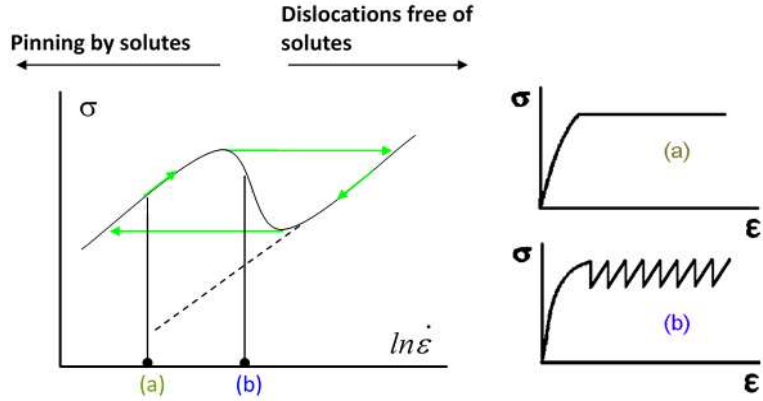

***Figure 3****. Left: Scheme explaining transformation of a monotonously increasing $\sigma(\ln \dot{\varepsilon})$ dependence obeying the Arrhenius law for thermally activated motion of dislocations (indicated by a dashed line) into an N-shaped dependence due to efficient pinning of dislocations by solute atoms at low enough $\dot{\varepsilon}$. Arrows trace a cyclic motion for the case (b). Right: Schematic deformation curves corresponding to $\dot{\varepsilon}_a$ either outside the negative SRS region or within it, as indicated by the corresponding letters.*

The resulting deformation curves are schematically (omitting any work hardening) presented on the right-hand part of Figure 3. Namely, if $\dot{\varepsilon}_a$ finds itself in the negative SRS region (case $b$ in the figure), the evolution of the state of the deforming material in $\sigma - \dot{\varepsilon}$ coordinates will be presented by a cyclic motion between the left (slow) branch and the right (fast) branch of the $N$-curve, which will be translated into a serrated $\sigma(\varepsilon)$ curve shown in the right chart. A smooth deformation curve will be observed if $\dot{\varepsilon}_a$ is taken outside this interval (case $a$). It can be said in an approximate manner that the properties of the SRS function determine the strain-rate domain of instability for a given temperature.

Were the motion of all dislocations identical, serration patterns would be qualitatively similar to such periodic relaxation oscillations, albeit evolving with the work hardening of a material, as predicted by "local" models of the instability [4,34,35]. The inevitable inhomogeneity of plastic flow leads to complex behaviors in real materials (cf., Figure 1). Since the discovery of the PLC effect, several generic types of behavior in tension conditions have been determined experimentally [4,22-25, 40-43]. Figure 1 illustrates three major types of stress fluctuations observed in polycrystalline alloys, showing a sequence of transitions from type $A$ to $B$ to $C$, taking place when $\dot{\varepsilon}_a$ is decreased from the upper to the lower boundary of instability. A similar sequence of serration types is observed when the temperature is increased at a given $\dot{\varepsilon}_a$. Besides the shapes of the serrations, the transitions between the different types of the jerky flow are characterized by qualitative changes in the PLC band kinematics. The signatures of type $A$ serrations are the observation of periodic stress increases followed by backward drops to the nominal stress level, as well as a quasi-continuous propagation of deformation bands along the specimen's tensile axis (Figure 4) [44]. Each stress rise precedes the nucleation of a band that usually occurs near one specimen end. The subsequent propagation towards the opposite end proceeds at a lower stress and is usually accompanied by irregular stress fluctuations. When $\dot{\varepsilon}_a$ is reduced, more regular stress oscillations with an apparent characteristic scale are observed (type $B$). These are related to a chained nucleation of deformation bands in the neighboring sections of the specimen (Figure 5) [45]. Each stress drop can be put into correspondence with an individual band. Although type $B$ bands either do not propagate or move over short distances, their correlated occurrence is reflected in the common terms of "relay-race" or "hopping" propagation. At the lowest strain rate, deep drops are observed below the nominal stress level (type $C$ behavior). Each of these events is also caused by a separate deformation band. However, unlike the type $B$ case, it is not clearly correlated with the previous bands [46]. The lower $\dot{\varepsilon}_a$ is, the more random the band nucleation becomes.

The existence of a nomenclature of types is itself evidence of a nonrandom nature of the dislocation dynamics. However, phenomenological models based on the scheme of Figure 3 and not considering the intrinsic strain heterogeneity, explain neither the great variety of deformation curves nor the variation of the band kinematics with experimental conditions. Moreover, the complexity of irregular curves goes far beyond their classification into "types". In particular, additional $D$ and $E$ types are sometimes distinguished in order to include some specific patterns observed in commercial alloys [22]. Various quantitative methods of characterization of the complexity of unstable plastic flow have been proposed lately. This Chapter will mostly present the progress provided by the statistical and multifractal approaches. For the reader's convenience, references to papers developing other concepts will be provided.

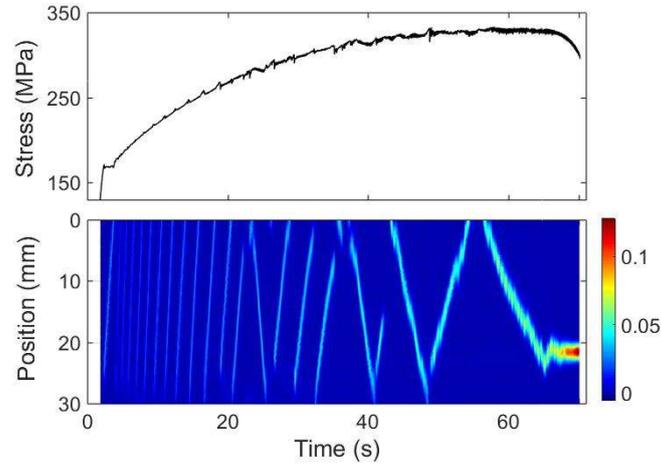

***Figure 4***. *Top: Example of a stress-time curve displaying type A serrations in an Al5.5%Mg alloy with Al₃Zr precipitates at $5 \times 10^{-3} \, s^{-1}$ [44]. Bottom: the corresponding local strain-rate map $\dot{\varepsilon}(x,t)$ showing propagation of deformation bands. The color bar represents the local $\dot{\varepsilon}$ scale in $s^{-1}$. The propagating strain-localization bands show up as bright oblique lines. Their vertical cross-section gives a rough estimate of the band width and the line inclination renders the band velocity. One can also discern a progressive transition to type B behavior at large strain (after roughly the $30^{th}$ second of the test).*

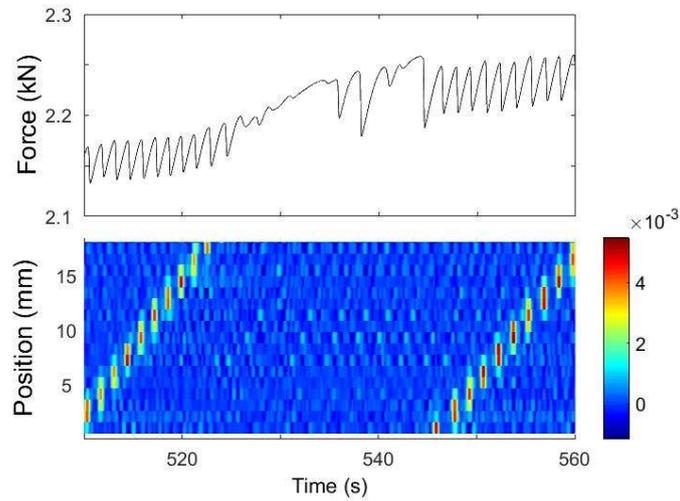

***Figure 5***. *Representation similar to Figure 4 for an Al3%Mg alloy deformed at $\dot{\varepsilon}_a = 2 \times 10^{-4} \, s^{-1}$, illustrating hopping propagation of deformation bands [45].*

The existence of a nomenclature of types is itself evidence of a nonrandom nature of the dislocation dynamics. However, phenomenological models based on the scheme of Figure 3 and not considering the intrinsic strain heterogeneity, explain neither the great variety of deformation curves nor the variation of the band kinematics with experimental conditions. Moreover, the complexity of irregular curves goes far beyond their classification into "types". In particular, additional *D* and *E* types are sometimes distinguished in order to include some specific patterns observed in commercial alloys [22]. Various quantitative methods of characterization of the complexity of unstable plastic flow have been proposed lately. This Chapter will mostly present the progress provided by the statistical and multifractal approaches. For the reader's convenience, references to papers developing other concepts will be provided.

## 2. Macroscopic scale

### 2.1. Statistics of stress serrations

Although the calculation of histograms of data distributions is straightforward, some common precautions need to be specified before presenting examples of research. A general problem concerning any of the analyses presented in this Chapter arises from an increase in the deforming stress because of the material work hardening (see Figure 1). The non-stationary character of the processed signal can lead to the wrong results of analysis [47]. Thus, the minimum required pretreatment of experimental data consists of subtracting the corresponding systematic trend evaluated using either a running-average or a polynomial fit $\overline{\sigma(t)}$ [48]. Moreover, as can be seen in Figure 1, the size $\Delta\sigma$ of stress drops may also increase on average during deformation, which would bias the corresponding statistical distributions. One way to avoid this pitfall is to perform calculations in time intervals where this trend can be neglected. As this approach restricts the statistical sample, a physically based normalization procedure is needed, which would allow one to deal with sufficiently long data series. In most cases, the slow trend can be removed by normalizing the deformation curve with respect to the average trend, $s(t) = \sigma(t)/\overline{\sigma(t)}$. The feasibility of such a procedure means that the evolution of the stress-drop size is mainly due to work hardening. Despite this obvious mechanism, the relationship between $\sigma$ and $\Delta\sigma$ is sometimes less straightforward. However, reconstruction of a stationary signal is usually possible by virtue of slightly more complex procedures, e.g., using a normalization function, $\Delta\sigma(t)$, found by fitting the evolution of stress drop amplitudes [48]. Examples of signals obtained after removal of the non-stationary course are illustrated in Figure 6 for three types of behavior of the PLC effect. Then, histograms of either normalized stress drops $\Delta s$ or peaks of the derivative of the normalized curve (right-hand column in Figure 6) may be calculated to characterize the statistics of the instability [49-51].

To interpret the following analysis, it is also useful to clarify that the statistical graphs will present results of calculation for a dimensionless variable $s = \Delta s / <\Delta s>$ rescaled by the average values of $\Delta s$ [52,53]. While such an additional pre-processing does not affect the histogram shape, this reduction makes it possible not only to compare statistical distributions for the different parameters of the stress-drops, e.g., amplitudes and durations, but also to include into this comparison physical quantities that characterize the deformation processes on distinct scales, e.g., the acoustic emission or bursts in local strain rates. Besides, such a process allows one to avoid arbitrariness in the choice of the bin size by using a unique bin in all cases. Examples of the so-constructed histograms for a binary Al3%Mg alloy are plotted in Figure 7 [49,54-58].

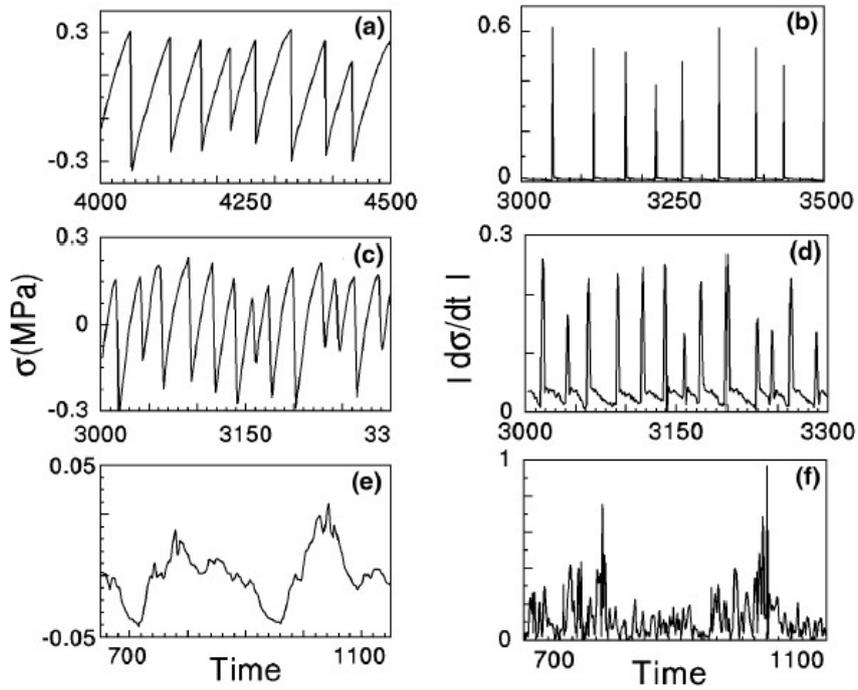

**Figure 6**. *Examples of time series obtained after removal of the non-stationary trend. Left: stress-time curves; Right: absolute value of the time derivative of stress. The strain rate value is increased from top to bottom: (a)-(b) $\dot{\varepsilon}_a = 5.56 \times 10^{-6} s^{-1}$, (c)-(d) $\dot{\varepsilon}_a = 2.78 \times 10^{-4} s^{-1}$, and (e)-(f) $\dot{\varepsilon}_a = 5.56 \times 10^{-3} s^{-1}$ (Figure from Ref. [51]).*

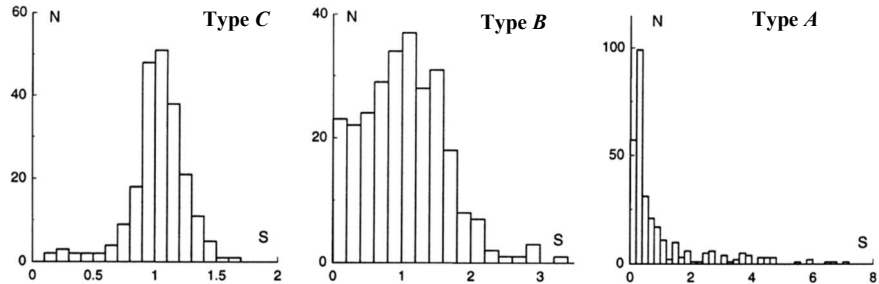

**Figure 7**. *Examples of histograms of the dimensionless stress drop amplitude s for three types of behavior of the PLC effect in a polycrystalline Al3%Mg alloy* [49,54-58].

Already these early attempts of analysis testified that the statistical distributions are qualitatively different for distinct types of behavior. The persistence of these features found for single crystals and polycrystals with different grain sizes and different chemical compositions indicated that statistical distributions may provide a quantitative characteristic distinguishing various behavior of the PLC effect. Therewith, while more or less complex-shaped peaks appearing for type *C* and *B* serrations indicate the presence of intrinsic scales and, therefore, make one think of random fluctuations about the ideal relaxation oscillations (it will become clear later how far it is from being true), a particular attention is attracted to a monotonously descending probability for type *A* behavior that does not reveal any characteristic scale.

Let us first consider the latter case. To characterize the statistics quantitatively, the probability for an event to have an amplitude *s* is calculated by counting the fraction $N(s)/N_{tot}$ of events within intervals $s \pm \delta s/2$, where $N_{tot}$ is the total number of events in the dataset and $\delta s$ is the bin size. It is readily noticed that large events are quite rare and many bins are empty. For this reason, a variable bin size is used to calculate the probability density function (PDF). Namely, $\delta s$ is taken constant in the intervals rich of events but increased in deprived regions until gathering a meaningful number of events (at least, five). Accordingly, the PDF is calculated to consider the bin variation [52,53]:

$$PDF(s) = \frac{N(s)}{N_{tot}\delta s}.$$ (1)

It occurs that type *A* serrations are often characterized by power-law statistics [49-58]. This behavior was established with certainty for stress drop amplitudes that are usually measured in a large dynamic range. The same conclusion is less reliable for their durations because the time resolution of load cells is rarely better than several milliseconds, which is similar to the time of the deformation band development [42,59-64]. Besides, the data may be biased by the reaction time of the "machine-sample" system (~ 0.1 s). Nevertheless, the data obtained testify that both amplitudes and durations of stress drops obey power-law dependences, as illustrated in Figure 8.

More exact methods based on the maximum-likelihood estimation with goodness-of-fit tests have been developed during the last decade to better handle the poor statistics of rare large events and detect power-law distributions in empirical data [65,66]. Nevertheless, the comparison of power-law exponents estimated by different methods showed that direct calculation with varied-bin correction renders satisfactory results for the PLC effect, in the sense that deviations from the values obtained by exact methods do not exceed the experimental uncertainty. Hereinafter, the above presentation of statistics will be used due to its intuitively clear interpretation and the opportunity of visual comparison of results obtained for distinct quantities.

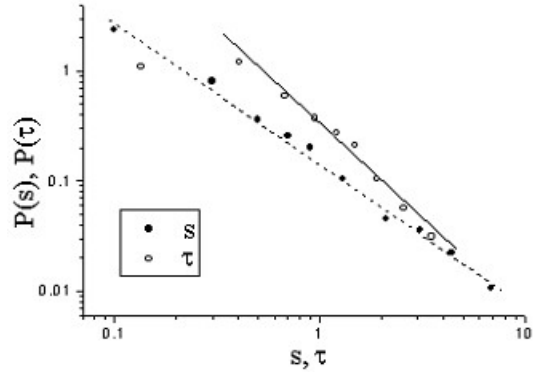

***Figure 8***. *Examples of normalized PDF for dimensionless amplitudes* $s$ *and durations* $\tau$ *of stress drops recorded at* $2 \times 10^{-3}$ $s^{-1}$ *in an Al3%Mg alloy.*

The observation of power-law dependences, which mathematically reflect the absence of a characteristic scale [indeed, $(kx)^a \sim x^a$], led to a suggestion that type $A$ behavior is governed by the mechanism of SOC. This concept was suggested to explain the abundance of scale-invariant behaviors in nature, including power spectral density of the "$1/f$"-noise, power-law statistics of avalanche-like processes, often referred to as crackling noise, formation of self-similar spatial structures, e.g., in earthquakes, forest fires, road traffic, and so on [67-70]. According to the SOC theory, the mechanism of scale invariance stems from the property that large complex systems naturally (without fine tuning of the order parameter) evolve to "a critical state in which minor events cause chain reactions of many sizes" [69]. The relevance of this concept for many natural phenomena is still a matter of debate, while many other models generating power-laws have been proposed [71-73]. More specifically for plastic deformation, various theoretical approaches can be found in recent reviews and original papers [74,75].

In spite of these debates, the SOC concept remains the hypothesis most frequently used to explain power-law statistics of intermittent deformation processes (e.g., [11,12,15,50-58,76]. It is supported by several experimental findings, most of which have been provided when finer scales were assessed by virtue of higher-resolution techniques, as will be discussed in Sec. 3. The following evidence in favor of this hypothesis is brought about by the analysis of the deformation curves themselves. It was predicted theoretically that in the case of SOC, the power-law exponents describing the statistical distributions of amplitudes and durations and the corresponding power-law relationship between these quantities have to be conform to power-law behavior of the Fourier spectrum, $S(f)$, of the deformation curve [77]. Using designations

$$PDF(s) \propto s^{-\beta}, \quad PDF(\tau) \propto \tau^{\gamma},$$
$$s \propto \tau^{h}, \qquad S(f) \propto f^{-\omega}, \tag{2}$$

where the scaling laws for the amplitudes and durations impose a relationship $h(\beta-1) = \gamma-1$, the following constraints are imposed on the spectral dependence: $\omega = 2$ for $2/h + \beta < 3$, otherwise $\omega = h(3 - \beta)$. In spite of the above-mentioned deficiencies in the quantitative examination of temporal behaviors, these relationships have been confirmed experimentally (Figure 9) [58]. For example, the $1/f^2$ spectrum shown in Figure 9b was obtained for a deformation curve characterized by exponents $\beta \approx 1.25$, $\alpha \approx 1.6$, and $h \approx 1.5$, verifying the condition $2/h + \beta \approx 2.6 < 3$.

At the same time, the transition from scale invariance at fast loading to peaked distributions at slower deformation (Figure 1) contradicts the SOC hypothesis. Namely, SOC models require loading at a vanishing rate, in order to assure independent nucleation of avalanches. On the one hand, this requirement seems to be satisfied for all $\dot{\varepsilon}_a$ used in experiments. Indeed, even the strain rate values about $10^{-3}$ s$^{-1}$ typical of type $A$ behavior still correspond to the conditions of quasi-static tests and may be considered as slow driving with regard to the values of about $10^2$ s$^{-1}$ - $10^3$ s$^{-1}$ on the right branch of the $N$-function (see Figure 3). On the other hand, the transition to peaked distributions with a decrease in $\dot{\varepsilon}_a$ raises serious concerns. For this reason, an alternative interpretation of scale-free statistics was proposed in terms of turbulent flow in dislocation ensembles [78,79]. Nevertheless, as will be discussed in Sec.3., investigations of deformation processes relevant to mesoscopic scales support the SOC hypothesis and, at the same time, provide a simple explanation of the transitions between distinct dynamics on the macroscale of stress serrations.

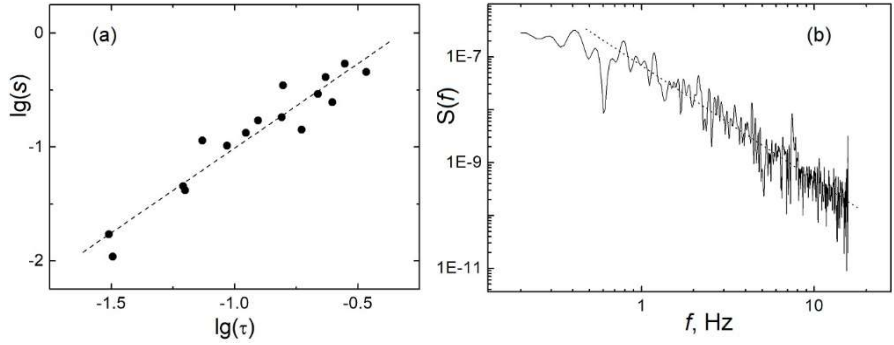

**Figure 9.** (a) Example of relationship between amplitudes and durations of type A stress serrations in an Al4.5%Mg alloy. The data are averaged for close τ values; (b) Fourier spectrum S(f) of the corresponding deformation curve. The dotted line traces the slope ω = 2.

## 2.2. Multifractal analysis

The understanding of scale transitions will however be incomplete without going deeper into analysis on the macroscopic level. As mentioned above, considering the transition to peaked histograms of stress drops and the degrading correlation between deformation bands with decreasing $\dot{\varepsilon}_a$, it is tempting to suggest that the dislocation dynamics approaches the ideal conditions of relaxation oscillations schematized in Figure 3. This limit would correspond to histograms in the form of the Dirac δ function, so that at first sight, random fluctuations about periodic behavior is a plausible explanation of the shape of real histograms. However, various analyses testified that such a situation is never reached experimentally, and the stress serrations do not correspond to random behavior even at the lowest strain rate of $\dot{\varepsilon}_a$ ~ $10^{-6}$ s$^{-1}$ attained in the experiments [46]. The general character of this conclusion for all strain rates was proved by virtue of multifractal (MF) analysis. Furthermore, an abundant literature treats this mathematical method [80-83]. The notion of fractal dimension was introduced by B. Mandelbrot to characterize scaling properties of self-similar natural objects [84,85]. The concept was later extended to heterogeneous patterns and signals which description requires multiple fractal dimensions [86,87]. A comprehensive description of the application of the MF method to stress-strain curves may be found in Ref. [48]. Below, only some basic notions that are necessary for the understanding of its practical application to the analysis of deformation processes will be presented.

Figure 10 explains the meaning of MF analysis by applying it to a self-similar (bottom sequence of peaks) and noise (top curve) signals. The formulae presented below consider discrete time series $\psi_k$ to mimic computer-assisted experiments ($k$ enumerates the data points). In order to reveal scale invariance, the interval representing the test duration is covered by a grid with a step $\delta t$. A local probabilistic measure $\mu_i(\delta t)$ is defined to characterize the local signal intensity in the $i$th interval. Its evident definition for discrete series is to calculate the summary signal within the box, normalized by the sum over all $N$ boxes:

$$\mu_i(\delta t) = \frac{\Sigma_i \psi_k}{\Sigma_N \psi_k}. \tag{3}$$

The next step is to construct partition functions, $Z_q(\delta t)$:

$$Z_q(\delta t) = \sum_i^N \mu_i^q(\delta t), \tag{4}$$

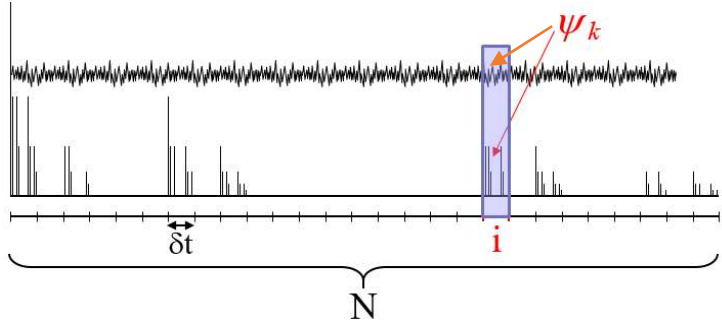

**Figure 10.** *Scheme illustrating calculation of a probabilistic measure (see Eq. 3) for time series $\psi_k$ defined over a grid with step $\delta t$. Two examples of time series (top: noise, bottom: sequence of peaks) are traced.*

where $q$ is a real number. It is easy to notice that by varying $q$, one makes dominate different $\mu_i$-values, i.e., different subsets of the signal, the feature known as a "mathematical microscope". Therewith, the subsets corresponding to a certain value of $\mu$ may have complex structure and be composed of boxes from different parts of the signal. The variation of $\delta t$ allows to assess scaling properties of $Z_q$ for a given $q$. Thus, the variation of both $q$ and $\delta t$ makes it possible to characterize scaling in a complex heterogeneous object. It is easy to calculate $Z_q(\delta t)$ for a uniform signal, e.g., for a constant function or a signal constant on average. The latter may be represented by periodic or random series, for $\delta t$ large enough with regard to the characteristic period of the signal variations. In this case, the local measure has the same value for all boxes (see Eq. 3), equal to $1/N \propto \delta t$. Therefore,

$$Z_q(\delta t) = N\delta t^q \propto 1/\delta t \times \delta t^q = \delta t^{q-1}. \tag{5}$$

Thus, for any $q$ these dependences lie on a master linear curve with a unit slope in coordinates $\log Z_q/(q-1)$ *vs* $\log \delta t$ (see any of the cited works to include the particular case $q = 1$). Figure 11 illustrates such a trivial scale invariance for a random signal.

The situation is qualitatively different for a fractal or multifractal object possessing the property of self-similarity. In the former case, the signal is characterized by a unique slope < 1, which defines its fractal dimension, $D$. In the latter case, $Z_q(\delta t)$ dependences exist as well, but the unique scaling law is replaced with

$$Z_q(\delta t) = \delta t^{(q-1)D_q}, \qquad (6)$$

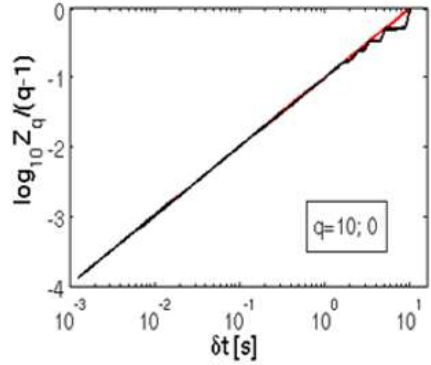

***Figure 11.*** *Results of calculation of partition functions (Eq. 4) for a random time series for two values of q.*

where $D_q$ values are called generalized fractal dimensions. Such a multifractal signal results in a fan of straight lines with different slopes, as illustrated in Figure 12 for a real deformation curve of an AlMg alloy. More exactly, as the time derivative of $s(t)$ highlights bursts of plastic activity, the analysis is applied to a time series obtained by taking the absolute value of its finite difference approximant [50,51]. The observation of non-trivial self-similar behavior reveals the presence of long-term correlations between intermittent events. Moreover, the analysis of experimental $Z_q(\delta t)$ dependences makes it possible to characterize these correlations quantitatively by a continuous function representing the spectrum of generalized dimensions $D(q)$, also called multifractal spectrum (Figure 13).

The interpretation of $D_q$ values is not straightforward. However, besides uncovering the bare fact of the presence of correlations, it brings significant quantitative infor-mation when one needs to assess the changes in the correlation strength upon modifications of the experimental conditions. Some $D_q$ values correspond to well-known dimensions possessing a clear physical meaning [80]. For example, $D_0$ renders the box-counting dimension of the signal's geometrical support (boxes with nonzero data values). Indeed, taking $q = 0$ will make all nonzero members of $Z_q$ equal to one in Eq. 4, so that $Z_q$ will simply give the number of nonempty boxes. In other words, $D_0$ characterizes the filling of the time interval with data. $D_1$ corresponds to the so-called information dimension, and $D_2$ gives the correlation dimension.

It is also clear from the above description that the height of the "$D_q$ spectrum" may characterize the signal heterogeneity. This quantity occurred to be quite sensitive to the transitions between different types of behavior of the PLC effect and provided a unique framework to characterize the changes taking place over a wide strain-rate interval [50,51]. In particular, the MF analysis allowed to prove that type $C$ serrations, considered for a long time as being caused by randomly occurring deformation bands, also possess a correlated temporal structure [46]. A detailed description of the multifractality of jerky flow can be found in Ref. [48]. In the present Chapter, we shall be interested in extending it to fine scales assessed by virtue of the AE method and comparing the conclusions provided by the MF analysis on various scales, as will be presented in Sec. 3.

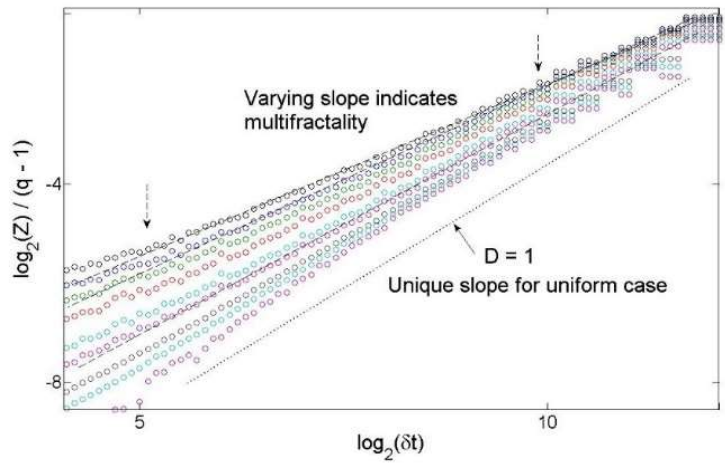

**Figure 12.** Partition functions (Eq. 4) for a deformation curve of an AlMg alloy. Different curves correspond to different q. Arrows indicate the scaling range.

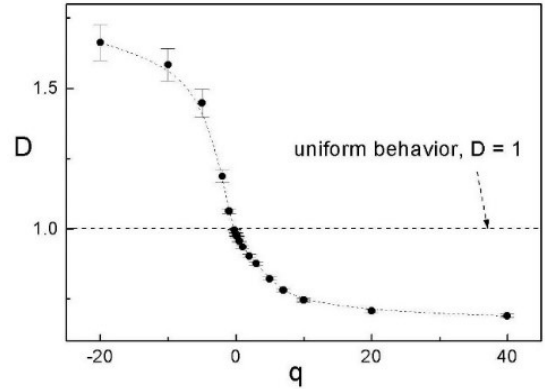

*Figure 13. Spectrum of generalized dimensions calculated for the data of Figure 12.*

## 2.3. Phase space reconstruction and other approaches

Besides the statistical and MF analyses, many other approaches to the investigation of serrated deformation curves have been suggested in the literature [88-99]. One result attracts a special attention in the scope of the Chapter, as it will assure continuity of the data interpretation. Let us notice that the transition from SOC to ideal relaxation oscillations means a crucial reduction of the number of degrees of freedom controlling the system dynamics. While the former characterizes systems with infinite dimensionality, the latter corresponds to a single degree of freedom, so that two observables, $\sigma$ and $\dot{\varepsilon}$, are sufficient to describe the dynamical state. The real behavior observed at intermediate and even low strain rates is much more sophisticated than this ideal situation. However, this complexity does not contradict the tendency to a reduction of the system dimensionality. Indeed, it is known that systems with a few degrees of freedom can perform very complex motions. This is the case of the deterministic, or dynamical, chaos that owes its name to an extremely high sensitivity of the phase trajectory, albeit deterministic, to initial conditions, so that the evolution of each variable makes one think of random processes [26]. Starting from the pioneering works in meteorology [100], dynamical chaos was detected in various natural and artificial nonlinear systems [26]. The possibility of chaotic dynamics in the dislocation system was predicted in the early 1980s [101]. Experimental investigations and numerical modeling started in the 1990s [50,51,88-90,101-103], at the same period as the statistical investigations of SOC-like behaviors. This synchronicity evidences once again the timeliness of the evolution of the plasticity theory to the analysis of collective behaviors of defects.

The problem can be illustrated as follows. Ideally, behavior of a system with a few degrees of freedom may be fully described by its trajectory in a $m$-dimensional phase space, so that the evolution of any variable can be traced (Figure 14, $a \rightarrow b$ operation). In practice, the evolution of only one or two variables is recorded experimentally in mechanical tests, e.g., the deforming stress and/or strain. It is therefore necessary to tackle the inverse problem (Figure 14, $b \rightarrow a$) and reconstruct the phase space with the dimension unknown *a priori*. The solution is rather direct in the case of linear systems, for example, with the help of the Fourier analysis. In contrast, behavior of a chaotic system can be very sophisticated because of a specific geometry of its attractor. A comprehensive description of the method of phase reconstruction, also referred to as dynamic analysis, goes beyond the scope of this Chapter and can be found in the literature [26,51,89,90]. It is important in the present context that the attractor is fractal: it appears the same on different scales and is therefore characterized by the property of self-similarity, hence its name "strange attractor". The considered approaches to nonlinear dynamics are thus fundamentally related [85].

Such processing revealed that the dynamics reflected in jerky flow corresponds to deterministic chaos in the conditions of type $B$ behavior, as illustrated in Figure 14. In particular, jerky flow in a Cu-Al single crystal was shown to correspond to the system dynamics resulting from non-linear interactions of only four modes [89]. A dimension of six of the reconstructed phase space was found for an AlMg polycrystal [50,51].

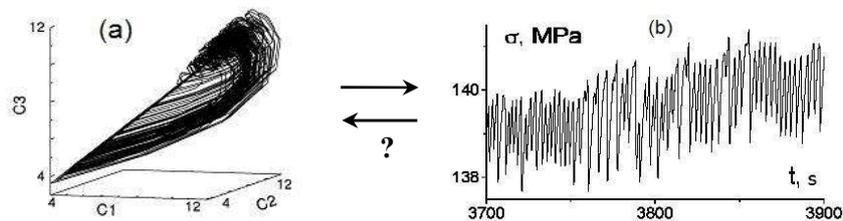

***Figure 14**. 3D projection of a strange attractor corresponding to dimension m=6 (a), which was reconstructed starting from a deformation curve shown in Chart (b). (for more detail on the coordinates in the phase space, see, e.g., [7]).*

The examples of the paragraphs *2.1 - 2.3* illustrate the need to explore various mathematical methods to characterize the entirety of experimental situations. For example, in contrast to type $A$, the statistics of serrations corresponding to chaotic behavior in type $B$ conditions is described by complex-shaped histograms revealing inherent scales (cf. Figure 7), often bimodal, which do not provide by themselves an interpretation of the behavior observed. Among diverse approaches, one can mention random-walk analysis [91,92], recurrence analysis [93], nonstationary spectral analyses including time-frequency methods (Cohen representation [94,95], wavelets [96]), Tsallis statistics [97], entropy-based methods [99-100], and so on. Without going into details of these methods, it can be stated that all authors conclude on nonrandom behaviors in all conditions reached in real experiments. Therewith, various kinds of behaviors correspond to different dynamical regimes. Importantly, a common property of different observations is the existence of scaling laws. Therefore, a more thorough comprehension of unstable plastic flow requires similar analyses with resolutions higher than that of mechanical tests, which will be considered in Sec. 3.

### 2.4. A possible dynamical mechanism

The results obtained due to the analysis of deformation curves allow to put forward a hypothesis on a dynamic mechanism that controls the heterogeneity of deformation during plastic instability. It suggests a dynamic equilibrium between the recurrent strain heterogeneity caused by the PLC bands and plastic relaxation of the resulting internal stresses during reloading after stress drops. This conjecture implies different characteristic scales: on the one hand, intrinsic scales (the plastic relaxation time, $t_R$, and a correlation length, $l_P$, over which the excess of internal stresses in the deforming region contributes to slip activation in other regions), and on the other hand, the reloading time, $t_L$, i.e., the scale imposed by the test conditions. In the ideal case of very low $\dot{\varepsilon}_a$ ($t_L \gg t_R$), which makes possible an efficient homogenization of local strains during reloading, this mechanism would lead to relaxation oscillations associated with random nucleation of PLC bands (type $C$ behavior). When $\dot{\varepsilon}_a$ is increased, $t_L$ becomes insufficient to fully relax local strain incompatibilities. Some spatial correlation of PLC bands sets up, and the periodicity of relaxation oscillations is disturbed. While $\dot{\varepsilon}_a$ is low enough, mechanical behavior has characteristic type $C$ properties, but the correlations lead to self-similarities that are manifested in the corresponding MF spectra. At the same time, the data analysis for similar samples tested in the same experimental conditions showed a strong variation in the resulting MF spectra, thus revealing a transient nature of this behavior [50,51,104,105]. When $\dot{\varepsilon}_a$ is increased, the gradual lessening of internal stress relaxation reinforces spatial correlation and leads to a relay-race propagation (type $B$), and then to quasi-continuous propagation (type $A$) of PLC bands. These changes in the spatial appearance are accompanied by changes in the shape of the deformation curves associated with particular dynamical regimes (deterministic chaos and SOC), as reflected by their statistical and multifractal properties. Such a qualitative consideration is corroborated by the observations that the deformation curves recorded in the transitory conditions, $C/B$ or $B/A$, correspond to large $D_q$ ranges revealing high heterogeneity levels, while pure types are characterized by relatively narrow MF spectra.

This scheme qualitatively explains the most studied case of tensile tests in a hard machine, but also the persistent propagation of deformation bands at any stress rate in a soft machine which does not allow for stress relaxation between instability events [106-109]. It may thus be expected that various experimental results could be described within the framework of models combining the microscopic property of negative SRS and mesoscopic aspects associated with the heterogeneity of plastic deformation and relaxation of internal stresses, controlled by the crystal structure and the defects microstructure. Although the need for consideration of the complex microstructures presents obvious difficulties, this approach has already provided a significant progress in the modelling of plastic instabilities [102,110,111].

## 3. Mesoscopic scale. Acoustic emission.

Investigations of plastic flow of solids with the help of the AE technique have a long history reported in numerous books and reviews [112-114]. The applications of this method are remarkably diverse. So, surveying average parameters for the acoustic activity and intensity, e.g., the average count rate or the cumulated amplitude, provides valuable information on the work-hardening stages, the intermittent processes of dislocation multiplication, the overall growth of the dislocation density, and the activity of different slip systems (e.g., [115-117]). Fine frequency and amplitude analyses are applied to distinguish distinct deformation mechanisms, such as dislocation glide, twinning, micro-cracking, and phase transformations [112-114]. A large number of studies began in the late 1990s, and were devoted to the statistics of AE that is generated during smooth plastic flow in the absence of macroscopic instabilities [74,118-121]. It occurred that even at macroscopically stable flow, the AE is not solely represented by continuous noise, as is expected for the case of uncorrelated motions in the dynamical system comprising billions of dislocations per $cm^2$. The AE also contains a discrete component manifested by short pulses with amplitudes that may exceed the continuous signal by orders of magnitude. An important conclusion of these works, realized on a great number of materials that ranged from ice to diverse metals and alloys, is that discrete pulses show ubiquitous power-law statistics. This finding led to a conjecture on an intrinsically avalanche-like nature of the dislocation dynamics on a mesoscopic scale, which is usually interpreted in terms of SOC as self-organization of dislocations towards a critical state. It is noteworthy that a similar conclusion was drawn from investigations of electric pulses caused by electron drag by mobile dislocations. Although these signals were very weak, they could be measured during unstable plastic flow of metals at liquid helium temperatures [122]. The corresponding statistical distributions of amplitudes and durations of such pulses also displayed power-law dependences in some range of variables, thus agreeing with the SOC hypothesis [123].

The comparison of this suggestion with the statistics of macroscopic stress serrations in ageing alloys puts forward several questions concerning the problem of micro-macro transition in plasticity. For example, can the above conclusion on the AE statistics be extended to include ageing alloys? In other words, does the AE obey power-law statistics in the conditions of the PLC effect or inherits the transitions from scale-free statistics to the distributions with a characteristic scale, as established for jerky curves? In other words, if the intermittency of plastic deformation is confined to fine scales but smoothed out on deformation curves for most materials, why does it develop into macroscopic jerkiness in the case of ageing alloys? These questions are addressed below with the aid of AE investigations of jerky flow.

### 3.1. AE recording and application of statistical analysis

The experimental methods used to obtain the results presented below were described in detail in [124,125]. In typical statistical experiments, the acoustic response to plastic flow is recorded using one piezoelectric transducer clamped to one surface of a flat tensile specimen near one of the grips. In the case of non-flat samples or compression tests, it is usually attached to a grip near the specimen end. Thus, the transducer gathers signals from various sources acting within the plastically deforming specimen, similar to collecting earthquake statistics on a seismic station. This approach is well justified for laboratory size flat samples composed of metallic materials which have low coefficients of attenuation of acoustic waves guided by the sample boundaries [126]. A special grease is used to assure a good acoustic contact. The most used transducers have a flat response in a frequency band about 0.1-1 MHz. The signal is pre-amplified and registered with a typical sampling rate of 2 MHz or 4 MHz. The latter choice follows the Nyquist criterion requiring the sampling rate to be at least twice as high as the highest frequency in the signal [127]. Such fast data acquisition results in huge data files. Accordingly, when only the statistics of acoustic pulses and not their exact waveforms are of interest, a software built in the acoustic systems allows for "real-time" extraction of meaningful acoustic events (hits) during the test. Figure 15 designates the parameters used for this purpose [24]. The event starting time, $t_0$, corresponds to the instant when the acoustic signal exceeds a threshold voltage $U_0$, set at a level depending on the noise measured in the free-running deformation machine (23 dB to 25 dB in the examples presented in this Chapter). The end of the event, $t_e$, is detected if the signal remains below $U_0$ longer than for a hit definition time (HDT). Afterwards, the system does not record hits during a hit lockout time (HLT), or a "dead time". The HDT and HLT allow to avoid recording as separate hits the unwanted echo signals caused by sound reflections from the specimen surfaces. As the sound velocity is high in metallic samples and their size is usually small, the echo return time is most often below 10 µs. For this reason, HDT is commonly taken large enough, e.g., 100 µs, to include all echoes into the event. Consequently, the HLT can be taken short, typically 20 µs – 50 µs, in order to avoid a significant loss of meaningful events. As a matter of fact, the amplitude AE statistics occurred to be quite robust regarding the choice of HDT and HLT in the range from 10 μs to 600 μs [128]. On the contrary, it is obvious that the measured value of the duration depends on the HDT and may include both the main event and secondary echoes. It is thus important to be careful when considering the duration statistics [129,130]. The further illustrations will be based on the amplitude statistics. To complete the technical details, it should be added that the peak amplitude is determined using a peak definition time (PDT), as the local maximum that has not been exceeded during PDT. This parameter allows one to avoid false peaks that may be caused by short sound propagations. The PDT is often taken to be equal to half of the HDT. It should be specified that the device stores the logarithmic amplitudes, $A_{log}$, of the events. To avoid confusion, the notation $A$ will be used for the peak amplitude after the conversion from logarithmic to linear units.

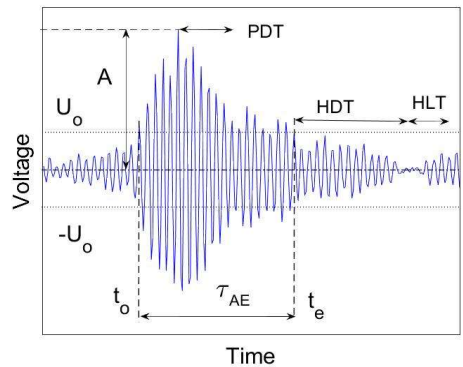

**Figure 15**. *Definition of the parameters used to extract individual AE events (see text for the definitions of the variables). (Figure adapted from Ref. [24]).*

The statistical analysis followed the same directions as those described above for stress serrations. To compare the statistics of the stress serrations and AE, let us notice that the stress drop, $\Delta\sigma$, reflects the mechanical work, $\sigma\Delta\varepsilon$, dissipated during the respective deformation process. Indeed, considering that $\Delta\sigma \ll \sigma$, the applied stress $\sigma$ may be taken to be approximately constant in this relationship. Furthermore, $\Delta\varepsilon$ is proportional to $\Delta\sigma$ because the latter is determined by the elastic reaction of the mechanical system: $\Delta\sigma = K\,\Delta\varepsilon$ ($K$ designates the stiffness value). Therefore, the statistics shown in Figures 7 and 8 represent the energy distribution of the plastic instability events. Accordingly, an adequate energy characteristic is searched for the AE analysis. However, the direct characteristic, i.e., the energy obtained by the integration of the acoustic event envelope (Figure 15), can suffer from the uncertainty caused by the possible joining of secondary echoes to the main hit. Moreover, it would be strongly sensitive to the ability of the recording system to separate events corresponding to distinct deformation processes but forming a dense sequence, e.g., because of subsequent triggering of dislocation avalanches [128]. For these reasons, the variable analyzed below is the squared amplitude $A^2$ which was argued to be proportional to the energy dissipated by the viscoplastic deformation giving rise to the AE event [119,131]. Similar to the variable $s$, histograms of normalized intensity, $I = A^2/<A^2>$, will be illustrated below.

Like stress serrations, AE amplitudes may also evolve during the test. Suggesting a physically based normalization procedure is not obvious in this case. However, the number of AE events is usually rather large, so that it is possible to calculate distributions over intervals where the AE is approximately stationary. Moreover, such a subdivision of the test interval allows for assessing the evolution of the distribution shapes over the course of the test.

### 3.2. AE amplitude statistics

Surprisingly, investigations using various Al-based alloys prone to PLC instability showed that the AE amplitude statistics obey power laws in all conditions, even for types $B$ and $C$ behaviors that are characterized by peaked histograms for the stress serrations [24,128-130]. For example, Figure 16 displays the PDF-dependences for AE collected before the onset of macroscopic instability in an AlMg alloy with two grain sizes differing by a factor of 2 (as-delivered and annealed conditions) [130]. Figure 17 presents similar dependences for AE recorded before and after $\varepsilon_{cr}$ in an AlMgScZr alloy [52].

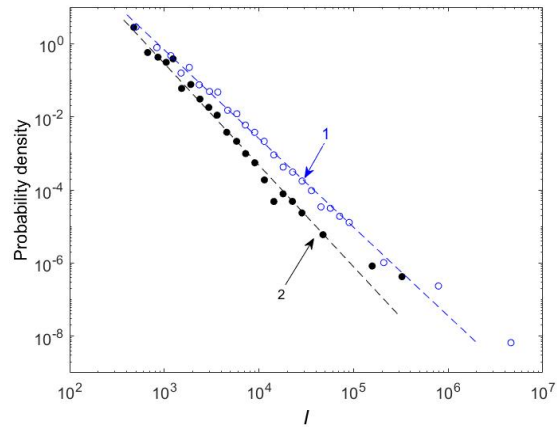

***Figure 16***. *PDF functions for AE intensity in a polycrystalline AlMg alloy. (1) As delivered specimen, $\beta_{AE} \approx 2.5$; (2) Annealed specimen, $\beta_{AE} \approx 2.9$. The grain size in the annealed state has increased by a factor of 2. $\dot{\varepsilon}_a = 2 \times 10^{-4}\,s^{-1}$ (Figure adapted from Ref. [130]).*

Thus, similar to stable plastic flow, dislocation processes appear to be avalanche-like on the scales relevant to AE. These observations allow one to specify the questions asked at the beginning of this Section. For instance, how can persistent power-law statistics of AE be compatible with peaked distributions found for type *B* and type *C* stress serrations? Furthermore, is type *A* behavior associated with a unique power law over a large-scale range including both the deformation curves and the accompanying AE or these scale levels are related to different statistics, which would be indicative of specific dynamical mechanisms? From the practical point of view, Figures 16 and 17 attract attention to a dependence of the power-law exponent $\beta_{AE}$ on the material microstructure. Calculations in different strain intervals show that it evolves over the course of the test [24,52]. Moreover, this evolution is not unique. The comparison of data for various materials reveals that its course may even change sign in alloys with different chemical compositions.

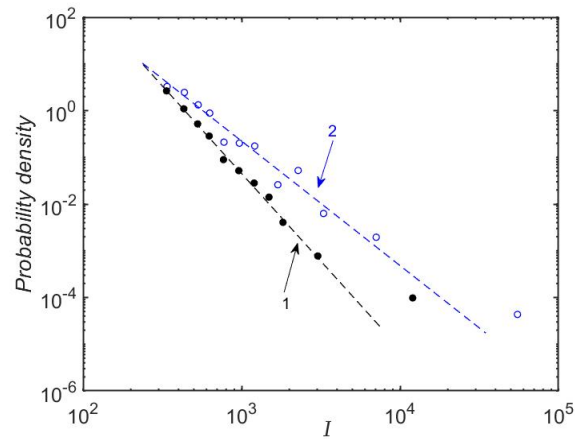

***Figure 17***. *PDF functions for AE intensity in a polycrystalline AlMgScZr alloy. (1) Before $\varepsilon_{cr}$, $\beta_{AE} \approx 3.6$; (2) After $\varepsilon_{cr}$, $\beta_{AE} \approx 2.4$. $\dot{\varepsilon}_a = 10^{-3}\,s^{-1}$ (Figure adapted from Ref. [52]).*

An answer to the first question can be found by surveying the AE generated at different instants of jerky flow. As the existence of a characteristic scale of stress drops is particularly pronounced for type-*C* behavior, Figure 18 shows a portion of a stress-time curve comprising two stress drops at a low strain rate, $\dot{\varepsilon}_a = 2 \times 10^{-5}\,s^{-1}$, and the accompanying AE signal [132]. The continuous signal appears as a black horizontal band including both the experimental noise and a possible contribution from uncorrelated motions of individual dislocations and/or small dislocation groups, e.g., dislocation pile-ups. The signal exceeding this "noise" displays a discrete series of individual AE hits seen as vertical bars on the time scale of the plot.

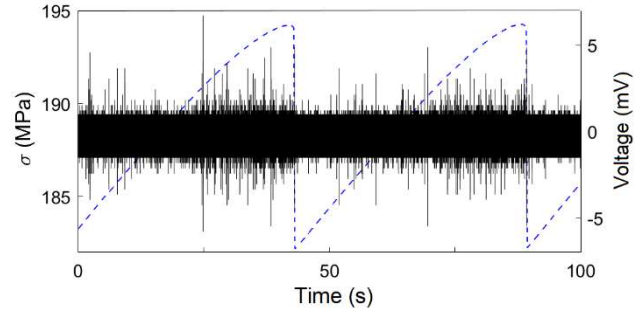

***Figure 18****. Example of AE signal in a time interval comprising two stress drops during tensile deformation of an Al3%Mg alloy at $\dot{\varepsilon}_a = 2 \times 10^{-5} \, s^{-1}$ (Figure adapted from Ref. [132]).*

Quite unexpectedly, the amplitude of the hits recorded at the instants of stress drops has nothing extraordinary regarding those occurring during the smooth reloading intervals. This similarity is well seen using a plot of AE amplitudes (Figure 19b). Instead, the stress drops are accompanied by bursts in the AE event duration $\tau_{AE}$ (Figure 19c) [24].

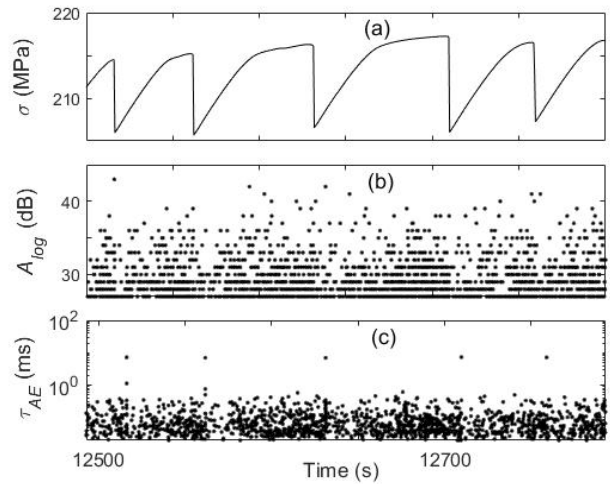

***Figure 19.*** *Portion of a serrated deformation curve of an AlMg alloy (a) and characteristics of the accompanying AE events: Logarithmic amplitude $A_{log}$ (b) and duration $\tau_{AE}$ (c). $\dot{\varepsilon}_a = 2 \times 10^{-5} \, s^{-1}$.*

The cause of such bursts is clarified in Figure 20 that compares typical waveforms of AE hits observed in different situations [130,133]. The intervals of smooth plastic flow are mostly accompanied by short, isolated hits with a rise time of several microseconds and duration of several tens of microseconds (Figure 20a). As argued in [134], their waveform is mainly determined by the properties of sound propagation in the material. Such individual events are sometimes observed during stress drops. More often, however, stress drops are accompanied by AE hits with complex shapes and durations varying from hundreds of microseconds to tens or even hundreds of milliseconds (Figure 20b). The data of Figures 18 to 20 thus lead to a conjecture that avalanche-like deformation processes are essentially the same at smooth and jerky flow, and the stress drops are not caused by extremely powerful dislocation avalanches but rather by consecutive triggering, or chaining of many dislocation avalanches in the same intensity range as during smooth deformation.

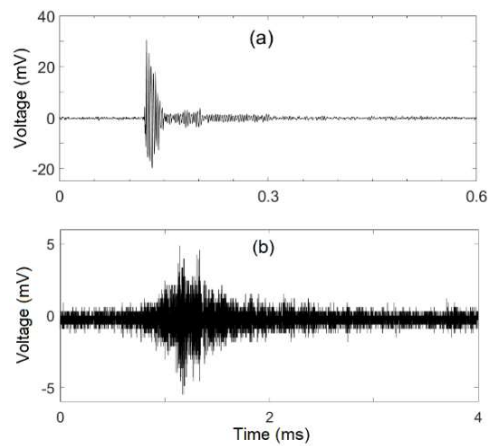

**Figure 20**. *Example of a short, isolated AE burst recorded at reloading after a stress drop (a) and a complex event accompanying a stress burst (b).*

A direct confirmation of this conjecture is provided due to a particular feature of plastic flow in the conditions of type *C* behavior. Although deep serrations occur abruptly after reaching the critical strain, the preceding deformation is neither smooth but displays lower-amplitude stress drops which often start occurring as early as during the elastoplastic transition and do not completely disappear beyond $\varepsilon_{cr}$ (Figure 21) [135]. These small drops were often attributed to sporadic fluctuations and disregarded in the literature on the PLC effect [136], but detailed investigations indicate that they are also caused by the DSA mechanism [135]. Their observation at the beginning of plastic flow allows for the visualization of clustering of AE hits, as can be seen in Figure 22 [24]. The difference between Figure 22 and Figure 19 can be easily observed. As the dislocation density is low during the early stages of work hardening, the chaining of deformation processes must be less pronounced than during established jerky flow. Accordingly, such small drops are accompanied by hit clusters that constitute dense sequences but can be individualized.

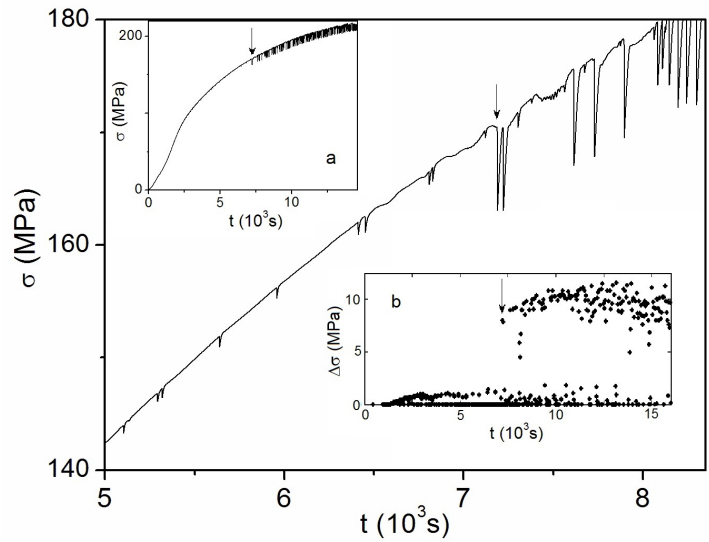

*Figure 21. Portion of a deformation curve $\sigma(t)$ around $\varepsilon_{cr}$ for an Al3%Mg alloy. Insets: (a) Global view of the deformation curve; (b) Amplitude of stress drops versus time. Arrows indicate the onset of type C stress serrations at $\varepsilon_{cr}$.*

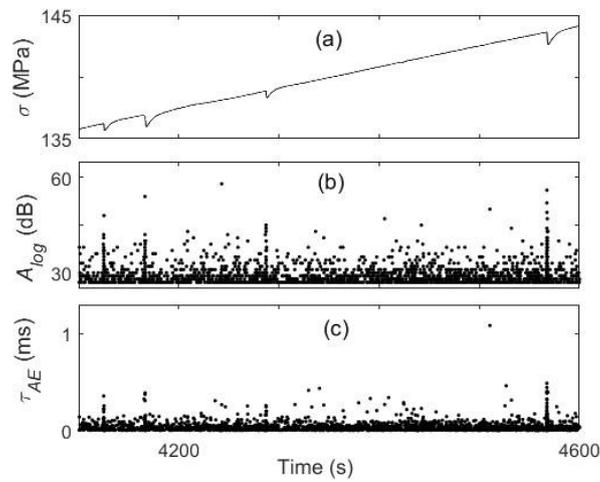

**Figure 22**. *Representation similar to Figure 19 for an early stage of deformation before the onset of the macroscopic instability in the same sample of an AlMg alloy. (a) – Stress-time curve; (b) – Logarithmic amplitude $A_{log}$ of AE events; (c) – Their duration $\tau_{AE}$. It is shown on a linear scale in order to better mark the events clustering at the instants of low-amplitude stress drops (Adapted from Ref. [24]).*

The clustering is progressively condensed when the material is work hardened so that groups of hits tend to degenerate into single events with very long duration, notably when the HDT is chosen relatively large (300 µs in Figures 19 and 21). This tendency also agrees with the obvious correlation of the frequency of hits with large stress drops: the activity is increased close to the stress-drop and reduced immediately after (Fig. 19, see also correlation analysis in [24]).

It may be noticed that the amplitudes and durations of the most intense AE hits recorded at small stress serrations are higher than the average level (Figure 22). This is not surprising because the AE is usually strong at the beginning of the test due to intense multiplication and large free paths of dislocations in the unhardened material [137,138]. Nevertheless, in this case either the maximum amplitudes are not higher than during perfectly smooth intervals. Finally, after some strain hardening, the stress drop events become completely indistinguishable with regard to the AE amplitudes that show a uniform scatter, while the respective duration bursts are re-inforced (Figure 19).

The overall pattern is illustrated by a cross-plot that displays the amplitudes and durations for the entire series of the AE events, as recorded during the test (Figure 23) [132]. The cross-plot reveals that the events are split into two groups. The upper group corresponds to the long hits detected during stress drops events. The lower group, on the other hand, unifies the other (more numerous) events gathered after $\varepsilon_{cr}$ with all the events recorded before $\varepsilon_{cr}$, thus confirming the conjecture of the same nature of dislocation avalanches during smooth and jerky flow.

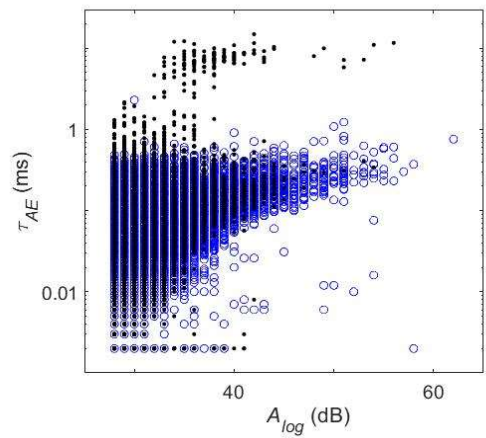

**Figure 23**. Examples of a $\tau$-$A_{log}$ cross-plot for the AE gathered before (circles) and after (dots) $\varepsilon_{cr}$. $\dot{\varepsilon}_a = 2 \times 10^{-5}\ s^{-1}$ (Figure adapted from Ref. [132]).

In [132], the conjecture about the chaining of dislocation avalanches was refined through the separate statistical analysis of subsets of AE events corresponding to smooth flow and stress serrations. Such an analysis allowed the authors to quantify a possible overlapping of AE events during the stress drops. Moreover, since in the case of type $C$ behavior the amplitudes of large and small stress drops are clearly separated by a gap of several MPa (see inset in Figure 21), it was also possible to separate such subsets. An example of this type of analysis is shown in Figure 24. It can be recognized that the PDF dependences closely coincide for smooth plastic flow, small serrations, and the entire dataset, and correspond to a power-law with $\beta_{AE} \approx 3.0 \pm 0.1$. The events recorded during the serrations reveal a crossover to a power-law with a shallower slope, $\beta_{AE} \approx 2.4 \pm 0.1$, in a range of higher amplitudes. It can be concluded that the deep stress serrations are characterized by an increase in the relative probability of high-energy AE events. As it is natural to suppose that the hits accompanying a stress drop are generated by dislocation avalanches from the same region corresponding to the deformation band, this trend reveals a possible superposition of AE hits due to the (quasi)simultaneous breakthrough of several avalanches. A somewhat elevated probability of strong AE events may even be suggested for other curves. This guess follows from the absence of a cut-off at relatively large energies, which is usually observed for various dynamical systems due to a finite system size which limits the avalanche size and also because of insufficient statistics of rare large events [74,75]. In any case, the fraction of overlapping events in the entire statistical sample is low and such a tendency does not noticeably bias the overall statistical behavior.

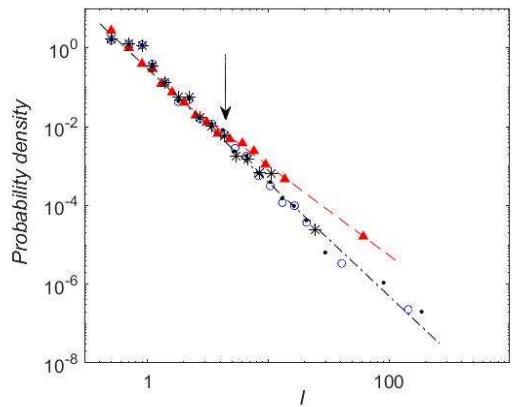

**Figure 24**. *Examples of PDF of normalized squared amplitudes I for different AE subsets: the entire set (dots), the hits recorded during smooth plastic flow (circles), during deep (triangles) and small (stars) stress serrations. Arrow indicates a crossover in the power-law dependence. The dashed-and-dot line has the slope $\beta_{AE} \approx 3.0 \pm 0.1$; the dashed (red) line corresponds to $\beta_{AE} \approx 2.4 \pm 0.1$, $\dot{\varepsilon}_a = 2 \times 10^{-5}\ s^{-1}$ (Figure adapted from Ref. [132]).*

Starting from the above description of the case of slow loading, the changes observed when $\dot{\varepsilon}_a$ is increased can be easily depicted. The increase in the plastic strain rate obviously leads to a global growth of the AE activity [24,130]. As a result, the clustering of AE events is globally enhanced, giving rise to some increase in the average τ value. However, the correlation between stress serrations and τ bursts, which reflects the clustering of dislocation avalanches, degrades progressively and becomes indiscernible in the fastest tests. Figure 25 presents these changes in a quantitative way. The separation of two data sets, as illustrated in Figure 23, becomes less pronounced with increasing $\dot{\varepsilon}_a$ and vanishes at the highest $\dot{\varepsilon}_a$ (Figure 25). At the same time, a clear power-law relationship occurs between $A_{log}$ and τ, similar to the relationship found for stress serrations (see Fig. 9). This regime of the PLC effect is characterized by ubiquitous power-law relationships for both AE events and stress serrations, in agreement with the application of the SOC concept to the high strain-rate dynamics.

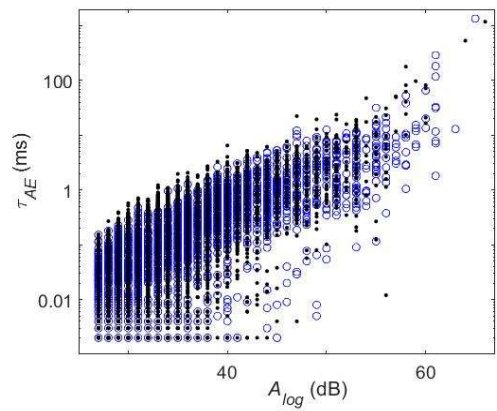

**Figure 25**. *The same plot as Figure 23 but for* $\dot{\varepsilon}_a = 6 \times 10^{-3}$ *s⁻¹ (Figure adapted from Ref. [132]).*

### 3.3. Further steps to analyze the dynamical mechanism

The data resulting from the statistical analysis of the AE that accompanies jerky flow connote several remarks:

The similar AE amplitude range during smooth deformation and at the stress drop events revokes the early opinion that large stress serrations correspond to intense discontinuous AE whereas smooth parts are accompanied by weak continuous AE. The questioned interpretation was based on the observation of AE count-rate bursts during stress drops [134]. However, as this discontinuity is a concomitant of $\tau_{AE}$ bursts, they have the same origin in the clustering of AE hits with ordinary intensity.

The deformation processes giving rise to long AE events at deep stress serrations should not be interpreted as elementary events in the sense of individual dislocation avalanches with large durations, but rather as packets of dislocation avalanches of various size. From the qualitative point of view, this statement follows from the observation of a progressive increase in the clustering strength with strain hardening (Figures 19 and 22). Besides, the separation of hits composing long events may be partly improved by varying the parameters used to extract individual events, e.g., by reducing the HDT. More quantitatively, it should be recalled that the avalanche amplitude and duration are related to each other by a power-law (Eq. 2), so that bursts in $\tau_{AE}$ should be accompanied by amplitude bursts, which is not the case. This remark is also consistent with recent observations of the formation of deformation bands with the aid of high-speed optical methods, albeit even the fastest of these tests had a much coarser time resolution (up to 5,000 frames per second) than the AE technique [61,62].

In spite of the salient property of scale invariance found for both stress serrations and AE, the apparent statistics and, therefore, the interpretation of the collective dislocation dynamics may depend on the surveyed quantity and the observation scale. This ambiguity is obviously related to the limitations of scale-free behaviors. The interpretation of the research results on the property of scale invariance in plasticity thus needs careful diligence. Although this warning is based on the above results for the PLC effect, it is particularly important for any investigation into the complexity of plastic flow in solids. As far as the PLC instability is concerned, we have already noticed that power-law statistics that are universally observed for AE are replaced with a progressive transition from power-law to peaked distributions of stress drop amplitudes. The separation between small and large scales may also occur for the same quantity, as demonstrated by the duration of acoustic hits for the type $C$ conditions. It was found that individual hits obey power-law statistics in a range of small durations (roughly, below 100 μs), in agreement with the SOC hypothesis, while bursts in $\tau_{AE}$ display a peak at large scale (> 1 ms). Moreover, such scale separation also characterizes stress serrations in type $C$ conditions. Even if the early statistical studies dealt with sufficiently deep stress drops and only displayed peaked distributions (Fig. 7), later measurements with a higher resolution allowed one to distinguish two scale ranges for small and large stress drops, as can be seen in Fig. 21. In view of the present discussion, it is not surprising that the statistical analysis of the amplitudes for small serrations led to a conclusion on power-law behavior (Fig. 26) [24]. As a whole, the distribution of stress drop amplitudes is bimodal at low $\dot{\varepsilon}_a$, although it is difficult to illustrate such a shape on the same plot because of the drastic scale separation. It is interesting in this connection that the distinction between small and large scales was also envisioned in seismology models. Even if the earthquakes are considered as a paradigm of SOC, deviations from the power-law in the form of a hump at large earthquake magnitudes were predicted as a consequence of the triggering of avalanches sequences, providing that the triggering avalanche is powerful enough to store sufficient elastic energy [139]. It is noteworthy that in contrast to laboratory experiments on plastic instability, the similar effect in seismology is difficult to observe experimentally because large earthquakes rarely occur.

It is also worth mentioning an additional argument in favor of the SOC associated with the PLC effect, which was obtained through analysis of the statistics of quiescent times between AE events [129]. Although the measurement of the waiting times is usually more certain than the measurement of durations that may be very short and biased by the recording system (see Section *3.1),* the waiting time statistics presented a puzzling question for many dynamical systems that are considered as candidates for SOC models. Since these models suggest a vanishing driving rate (see Section *2.1),* the statistics of the intervals between successive avalanches should obey a Poisson-like exponential law. However, power-law statistics of the waiting times were found experimentally for diverse systems including earthquakes, solar fluxes, and turbulent transport in magnetically confined plasma [139-143]. Various concepts were advanced to reconcile these observations with the SOC models, e.g., by attributing power-law correlations to a correlated driving signal [144] or to temporal variations of the activity rate [145]. In contrast, experimental studies of the PLC effect showed close-to-exponential behavior for the intervals between AE hits [129]. At the same time, a transition to power-law statistics took place when the low-amplitude component was cut using a threshold. This result corroborates a hypothesis that the apparent power-law behavior may have a general cause that is related to the cutting off of the apparent experimental noise [146-148].

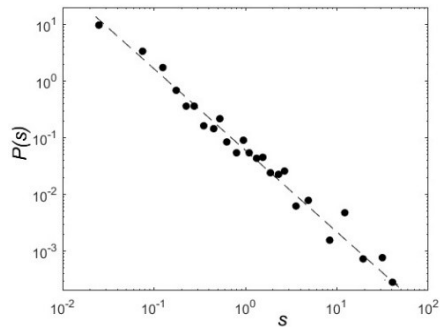

**Figure 26**. *Example of a normalized PDF for dimensionless amplitudes, s, of low-amplitude stress drops (Δσ < 5 MPa) recorded at a low strain rate, $\dot{\varepsilon}_a = 2 \times 10^{-5}$ s$^{-1}$, in an Al3%Mg alloy.*

The observation of a persistent power-law character of the AE amplitude statistics in the case of the PLC effect, on the one hand, and for numerous materials characterized by smooth plastic flow, on the other hand, allows for a unique approach to the question of intermittency of plastic deformation on a mesoscopic scale in such qualitatively different conditions. As a matter of fact, the problem of the meso-macro scale transition is even more vivid in the latter case because in view of the scale invariance established for dislocation avalanches, smooth deformation curves should not be observed. As discussed in [11], macroscopically stable deformation implies the existence of inherent factors confining the size of dislocation avalanches. In particular, such constraints may be caused by the intrinsic lengths that are related to the microstructure and to the crystallography of the dislocation glide. The results presented in this Section bear evidence that similar limitations must also apply to the dislocation avalanches in the conditions of macroscopic plastic instability.

Figures 16 and 17 bear witness that the power-law index may be indicative of changes in the material microstructure. This sensitivity to the microstructure agrees with the suggested dynamical mechanism considering that the conditions of internal stress relaxation play a preponderant role in the correlation of deformation processes (Sec. *2.4*). Therefore, it may be expected that the investigation of the AE statistics may bring quantitative information on the microstructure effect on the avalanche behavior of dislocations and, in particular, on the PLC instability. The literature data on $\beta_{AE}$ vary from 1.4 to 2 for smoothly deforming materials [121]. The lower values were detected for crystals with hexagonal lattices (ice, Cu, Zn, and Cd) in which plastic deformation is mostly constrained to one slip system. Steeper dependences obtained for cubic crystals (a value of $\beta_{AE} = 2$ was found for pure Al) were attributed to a stochastic factor caused by the multiple slip that led to forest hardening and the formation of dislocation structures. Indeed, being effective obstacles to the dislocation motion, these features may reduce the probability of large avalanches. Moreover, the obstruction to the self-organization of dislocations may reinforce the continuous uncorrelated AE and globally reduce its discrete component. The stochastic factor may be strengthened in alloys due to additional pinning of dislocations by impurity atoms in solid solution and also due to precipitates. For example, $\beta_{AE}$ values ranged from 2 to 3 and evolved during the deformation of binary polycrystalline AlMg alloys [24,130]. Even higher power-law indices (sometimes up to 4) were observed in AlMg-based alloys with precipitates [52]. Both a decrease and an increase in $\beta_{AE}$ were observed upon refinement of the grain structure. This uncertainty was attributed to possibly antipodal roles played by grain boundaries in different situations because they may serve both as sources and sinks of mobile dislocations, and also as obstacles to their motion [52,149].

To complete the description, it is also useful to recall that lower $\beta$ values, typically from 1 to less than 2, characterize stress serrations. This reduction is likely due to a relatively low sampling rate of typical load cells in deformation machines ($\geq 1$ ms), so that dense sequences of avalanches resolved by the acoustic system may appear as a single stress drop. Its amplitude will then be determined by the summary effect of many avalanches, thus increasing the probability of larger events and diminishing $\beta$. Thus, it may be suggested that the power-law statistics which characterize the dislocation avalanches that occur during DSA is correctly determined using the AE technique, whereas the exponents found for stress serrations may be underestimated.

In conclusion of this section, let us consider the AE data to further examine the dynamical mechanism, as proposed in *2.4* on the basis of analysis of stress serrations. It was argued above that the transitions between the different types of macroscopic behavior of the PLC effect can be explained in terms of the conditions governing the correlations between deformation bands. Let us now discuss if the suggested concept can also apply to correlations within individual deformation bands. We start again with the case of virtually uncorrelated bands of type *C*. For this case, it was conjectured that the effective homogenization of local strains during slow reloading after a stress drop destroys the memory about the previous strain localization, resulting in a loss of correlation between successive deformation bands. On the other hand, the same argument leads to the suggestion that there is a strong correlation between the deformation processes within one band. Indeed, as the material state becomes highly uniform prior to the next instability, various parts of the specimen reach the threshold stress (the maximum of the *N*-curve in Figure 3) nearly simultaneously. As a result, nucleation of a dislocation avalanche at some site may trigger a dense sequence of avalanches. The development of such a catastrophic strain burst will be stopped due to elastic unloading and a return back to the slow branch of the *N*-curve. Thus, the strength and duration of the deformation bands and the concomitant stress drops will be determined by the elastic properties of the machine-specimen system and the shape of the SRS function. Therefore, large instability events will have a typical size in conformity with the scenario of relaxation oscillations. As far as low-amplitude serrations are concerned, such events will occur when the instability threshold is reached in a less uniform part of the specimen, thus giving rise to fewer avalanches, without triggering a "catastrophic" process. It is also noteworthy that according to Figures 18, 19, and 22, the homogenization during macroscopically smooth loading takes place not only through the motion of individual dislocations and small dislocation pileups that generate continuous AE, but also through individual dislocation avalanches responsible for short discrete AE events.

It is clear in this framework why large driving rates are associated with SOC-type behavior. When $t_L \ll t_R$, the relaxation of internal stresses is insignificant. Besides, it is known that the *N*-curve is shallowed at high $\dot{\varepsilon}_a$ [4,34]. As a result, there constantly exist material elements close to the threshold of instability, so that the dislocation system finds itself in a globally critical state which allows for avalanches of any size.

The AE behavior observed for type *C* and, albeit less pronounced, for type *B* is analogous to the well-known phenomenon of synchronization in complex dynamical systems, which is manifested by the repetitive collective movements of either a part or the whole system composed of a large number of coupled oscillators [26]. A famous example of synchronization is demonstrated by the synchronous luminescence in firefly populations [150]. In the case of plastic deformation, the oscillatory character of the elementary processes is associated with the stick-slip character of the thermally activated motion of dislocations through obstacles. As far as the nature of coupling is concerned, several mechanisms can be envisaged, among which the coupling via elastic waves is considered to be predominant [12,74]. It can be stated that elastic waves are at the same time responsible for the studied phenomenon (i.e., the AE which carries information on the deformation processes) as well as important actors of the processes themselves. The analogy with the synchronization phenomenon is more than speculative. Indeed, a transition between SOC and synchronization has been predicted in some generic models [151]. In particular, such transitions and the coexistence of two dynamical modes are characteristic of the models that are based on block-and-spring chains. These types of models were also adapted in computer simulations of the PLC effect [49,152]. Two basic parameters control this behavior: the coupling strength and the nonlinearity of the driving force. In this scheme, SOC corresponds to a weak nonlinearity and strong coupling. Synchronization is found in the opposite case. It is clear from the above discussion that these criteria qualitatively apply to the PLC effect. Indeed, the shallowing of the $N$-curve at high $\dot{\varepsilon}_a$ corresponds to the weakening of the nonlinearity, while the lack of relaxation of the internal stresses is responsible for the strong spatial coupling.

### 3.4. What can be learnt from multifractal analysis of AE?

Similar to the case for the analysis of the jerky deformation curves, a further step to increase our understanding of collective dislocation dynamics may be provided by MF analysis. This type of analysis was conceived with the purpose of describing complex objects that can be characterized by the heterogeneous clustering of events or structures. Such investigations are rather sparse and are still at an incipient stage. Some results and questions raised by these studies are presented below.

Since the previous Section revealed that the AE signals stem from the same elementary processes (dislocation avalanches) during smooth and jerky flow, it is of interest to compare MF scaling for two kinds of time series, i.e., deformation curves and the accompanying AE signals. Examples of such a comparison for two strain rate values, which correspond to type $C$ and type $B$ behaviors, are shown in Figs. 27 and 28 [153]. The figures present families of partition functions (see Eq. 6) for time series represented by the stress-time derivative and by the amplitudes of AE events.

Without going into details, as discussed in [153], the following observations should be emphasized. Scaling in meaningful ranges that are spread over more than an order of magnitude of $\delta t$ are found in all cases, as shown by the straight lines traced in the plots. It is bounded from above by the length of the analyzed time interval and, in the opposite limit, by the minimum waiting time between the respective events (stress drops or AE hits). Significantly, the scaling interval for the AE signal covers that for the deformation curve and spreads to smaller scales without changing the slope when $\delta t$ becomes smaller than the minimum waiting time between stress drops, i.e., when it corresponds to the range pertaining to smooth deformation. It can thus be concluded that all AE events belong to the same MF ensemble, be they associated with either stress drops or smooth plastic flow. This result confirms the conjecture that AE events have the same nature over the entire deformation curve.

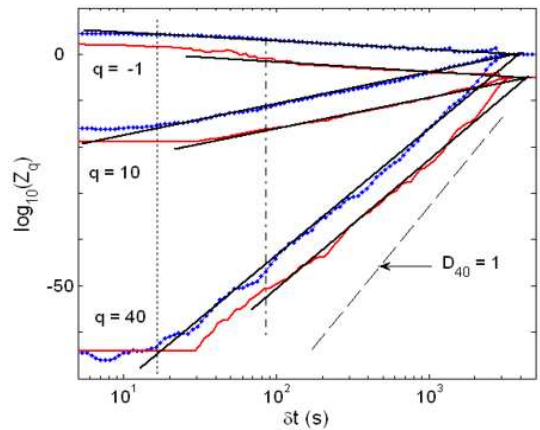

*Figure 27. Comparison of the partition functions $Z_q(\delta t)$ for AE time series (blue lines with symbols) and for stress-time series (red lines without symbols) for an AlMg specimen deformed at $\acute{\varepsilon}_a = 2 \times 10^{-5}\ s^{-1}$. The dependences for the stress-time data are shifted downwards to avoid superposition with their counterparts and facilitate the figure reading. The vertical dashed and dash-and-dotted lines indicate the lower scaling limit for AE and stress-time series, respectively. The straight line corresponding to the trivial scaling $D_q = 1$ is shown for the maximum $q$ value (Figure from Ref. [153]).*

Several refinements of this picture are worthwhile, namely: (i) In some cases, MF scaling is only found after the truncation of the smallest events using a threshold. This experience agrees with an intuitive suggestion that the least-intensive deformation events may not be a part of collective processes but occur at random. (ii) The scaling dependences may not have the same slope for the AE and the corresponding stress-time series. While the data of Fig. 27 present an example with similar $D_q$ values, as visualized in Fig. 29 with the aid of a MF spectrum, Fig. 28 demonstrates clear deflections at large enough $q$ values. These observations confirm the above statement that the apparent behavior may depend on the scale of observation, so that quantitative comparisons should be made with precaution. (iii) Reliable MF spectra were not obtained for series of AE hits at the highest strain rates (type *A* behavior), although the treatment showed clear tendencies of fractal scaling. This difficulty in detecting multifractality could be caused by the decreasing capacity of the AE technique to resolve individual AE events when the overall AE activity increases strongly. Nevertheless, this problem could be overcome by applying the analysis to as-recorded AE signals. Besides, this approach allowed for the examination of smaller-scale behavior at all strain rates, as presented in the following paragraphs.

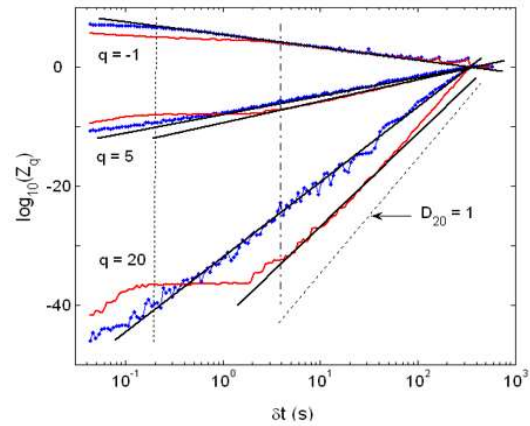

*Figure 28. The same as Figure 27, but for $\dot{\varepsilon}_a = 2 \times 10^{-4}\ s^{-1}$ (Figure from Ref. [153]).*

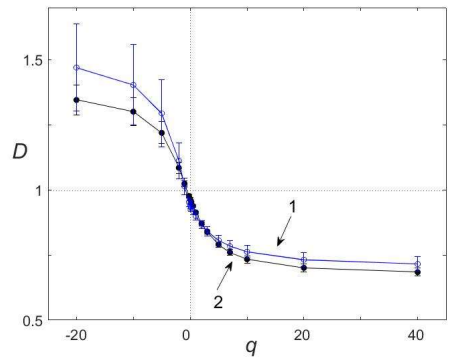

*Figure 29. Spectra of the generalized dimensions for the data from Figure 27. (1) stress-time series; (2) AE signal.*

The examples of Figures 27–29 dealt with MF analysis over long time intervals that contain many stress serrations. In such cases it was practical, at not very high strain rates, to extract discrete AE hits and process their series. However, this means that each hit is considered as reflecting an "elementary" deformation process and is reduced to one point in the time series, while the correlations characterizing behavior within a single stress drop and even over one period of "relaxation oscillations" escape from such treatment. For this reason, attempts at analyzing the continuously recorded AE signals, also known as the so-called datastreaming [154], were undertaken. Performing such treatment in intervals of different length allowed one to examine the spread of the scale invariance that characterizes the deformation processes occurring under various deformation conditions.

In the following examples, an alternative representation of multifractal behavior was applied in terms of the singularity spectra, $f(\alpha)$. Such spectra can be obtained via the Legendre transformation of the function $D(q)$: $\tau(q) = (q\text{-}1)D(q)$; $f(\alpha) = q\alpha\text{-}\tau(q)$; $\alpha = d\tau(q)/dq$ [80]. The practical method for calculating $f(\alpha)$ can be found in [86]. Although both representations are equivalent, the use of singularity spectra is useful because of its clear physical meaning. Namely, the singularity strength, $\alpha$, of the local measure describes its scaling with regard to the box size: $\mu_i(\delta t) \sim \delta t^{\alpha}$. The value of $f(\alpha)$ can be qualitatively defined as the fractal dimension of the subset of boxes corresponding to the singularity strength in a small interval around $\alpha$.

Figure 31 presents an example of such calculations for an AE signal recorded at $\dot{\varepsilon}_a = 2 \times 10^{-4}\,\text{s}^{-1}$, which corresponds to type $B$ behavior (Fig. 30) [125]. Results of the analysis for two intervals illustrate that a smooth MF spectrum is detected over an interval including one reloading/serration sequence, with a scaling range $\delta t \approx [40\ ms; 0.6\ s]$, and over an interval covering many stress serrations, with a scaling range $\delta t \approx [10\ \text{s}; 100\ \text{s}]$.

As can be observed, the two spectra are quite similar to each other. At the same time, the respective $\delta t$ ranges where scaling was found are not adjacent. Analysis of the time intervals with intermediate lengths showed a tendency to form a fan of partition functions, which allows one to expect that the break in scaling may be due to the truncation of the smallest events. In other words, it may be assumed that the same mechanism of correlation operates in a wide time scale range, from milliseconds to seconds. On the other hand, the break in scaling might also be related to the low AE activity after deep stress drops. It may be mentioned that in this context the MF analysis fails at large strains where the AE activity is strongly reduced, most probably because of the accumulation of obstacles to the dislocation motion. Therefore, the question regarding the spread of the correlations which give rise to multifractality needs further verification.

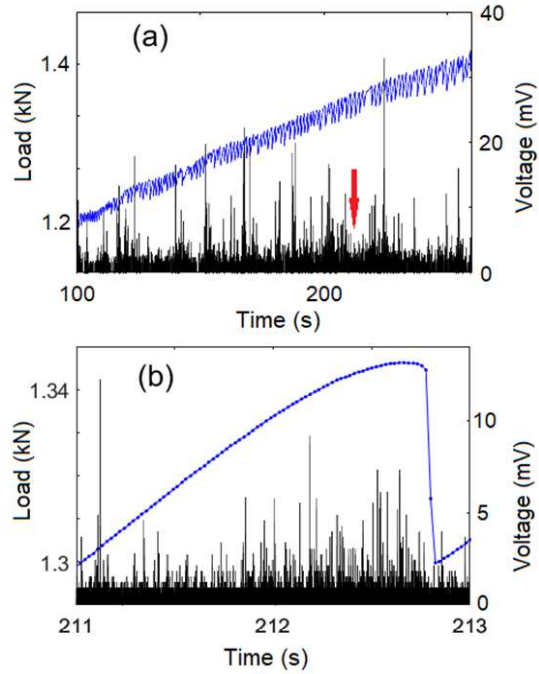

*Figure 30. Positive half of the AE signal (black color) accompanying type B serrations (blue color) in an AlMg polycrystal deformed at $\dot{\varepsilon}_a = 2 \times 10^{-4} s^{-1}$. (a) – Time interval covering a large number of serrations. The vertical arrow indicates the location of the short interval (b) that corresponds to one period of "relaxation oscillations", or one reloading/serration sequence.*

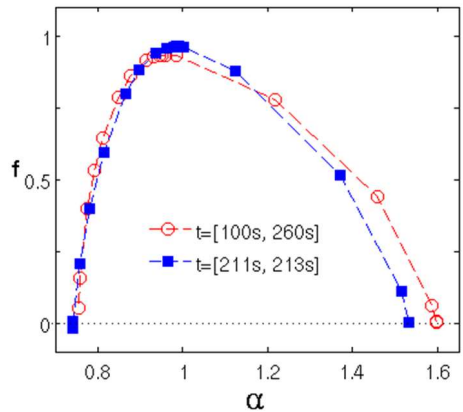

*Figure 31. Singularity spectra for the AE signal of Figure 30 calculated in the respective time intervals.*

Figures 32 and 33 extend this analysis to the scale of individual AE hits. Examples of single waveforms observed at $\dot{\varepsilon}_a = 2 \times 10^{-5}\,\text{s}^{-1}$ are shown in Fig. 32. As discussed above, deep stress drops are accompanied by complex signals with a millisecond duration (Figure 32a). The smooth reloading parts usually display unstructured short bursts with a short front that is followed by exponentially damped oscillations [130,133]. However, sequences of events were also observed that present interest for the analysis (Figure 32b). Smooth singularity spectra were found in both cases (Figure 33), with a scaling range from 1 ms down to several microseconds [125,133]. It is worth highlighting that thanks to the MF formalism, the existence of a fine structure within "elementary" instability events has been detected for the first time. Nevertheless, as the AE activity is relatively small at a low strain rate, the structured events are followed by significant periods which contain only short bursts or continuous noise. Consequently, an increase in the duration of the analyzed interval deteriorated and destroyed scaling behavior. Scaling was found again for long enough intervals containing many AE hits. Taking into account the above-said concerning the break in scaling on a time scale of seconds, it may be concluded that the AE may not be globally multifractal in the conditions of type $C$. Singularity spectra were found either for individual events (or their clusters) or for long enough series of events. Similar features were also observed at intermediate strain rates corresponding to type $B$ deformation curves. Such a break of multifractality agrees with the hypothesis of synchronization at low and intermediate strain rates, which implies a tendency to periodic behaviors.

A qualitatively different situation occurred under the conditions of type $A$ behavior at $\dot{\varepsilon}_a = 6 \times 10^{-3}\,\text{s}^{-1}$. As already specified above, AE signals fill the time axis quite densely at this strain rate due to a globally increased acoustic activity. In contrast to lower strain rates, multifractality was found for all time scales in this case. Such universality may indicate the formation of globally correlated behavior, as consistent with the conjecture of SOC at high strain rates.

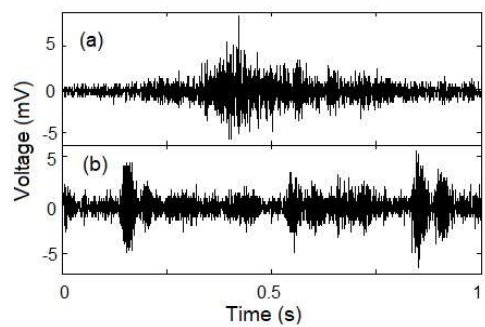

***Figure 32****. Examples of AE events observed during deformation of an AlMg sample at $\dot{\varepsilon}_a = 2 \times 10^{-5}\,s^{-1}$. (a) Typical signal accompanying a stress drop; (b) Sequence of signals sometimes observed during smooth deformation between stress serrations.*

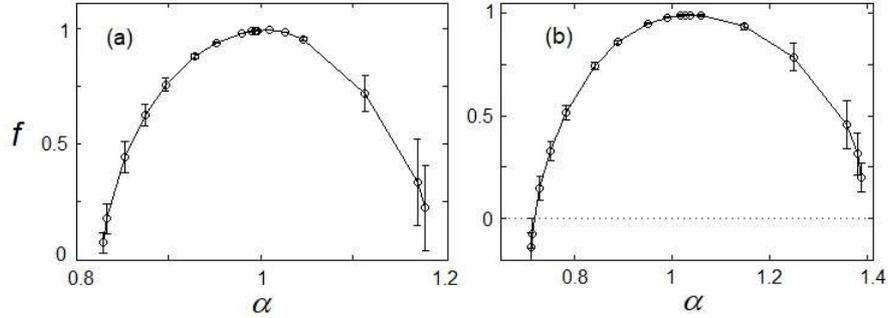

***Figure 33***. *Singularity spectra of AE events that are shown in Figure 32. The maximum f value, which corresponds to the fractal dimension of the support of the entire signal, is close to 1. Such a globally non-fractal geometry means that the AE completely fills the time interval.*

However, the results of this Section need a careful verification. Although scaling features were detected with certainty in the above examples, reliable quantitative determination of MF spectra and their comparison for different strain rates and different scale ranges was not possible systematically. Further investigations, perhaps, which use different methods of analysis are needed to better understand this phenomenon.

## 4. Conclusions and perspectives. Wave-intermittence duality.

### 4.1. Intermittence of plastic flow on multiple scales

The phenomenon of macroscopic plastic instability in dynamically strain ageing alloys presents various manifestations of the self-organization of crystal defects. In particular, the intermittent nature of AE allows for a conclusion on an inherently avalanche nature of deformation processes in a range of small scales. Taking into account the cited literature, this conclusion can be extended to all materials where the plasticity is governed by the motion of conventional crystal defects, *par excellence*, dislocations. The repartition between avalanches and uncorrelated movements of dislocations depends on the crystal structure, chemical composition, defect microstructure, and experimental conditions, e.g., strain rate, temperature, and sample geometry.

Although the well-known known manifestations of the PLC instability pertain to the macroscale, it involves the same elementary deformation processes as in the case of smooth plastic flow of any material. The basic elements of the collective dislocation motion on the mesoscopic scale are dislocation avalanches. The similar range of AE events during abrupt stress drops and smooth reloading intervals testifies that similar limitations of the avalanche size operate in both cases. At the same time, high $\beta_{AE}$ values, as compared with those reported for pure materials, indicate that DSA has an influence on the avalanche size over a significant portion of the deformation curve. Moreover, the macroscopic instability caused by the DSA engages additional dynamical mechanisms due to the nonlinearity of the SRS function. The occurrence of distinct kinds of macroscopic strain localizations and stress serrations is controlled by the conditions of correlation between dislocation avalanches. Although the avalanches themselves predetermine the ubiquity of scaling behaviors associated with the PLC effect, the variation of the conditions of their correlation is responsible for the diversity of manifestations of self-organization on the macroscale, such as SOC, chaos, and synchronization. In this context, the PLC effect attests itself as a unique object for laboratory investigation pertaining to the physically different realizations of complex dynamics on different scales for the same nonlinear system.

First attempts to extend the analysis of complexity to individual avalanches have brought evidence that such "elementary" events can also manifest self-similarity revealed by virtue of the MF formalism. This research deserves special attention because additional mechanisms of spatial coupling may acquire importance when the scale range approaches that of the individual dislocations, e.g., the mechanism of double cross-slip of dislocations. In particular, a crossover in the scaling dependences of the partition functions was observed at such scales in [125]. Theoretical arguments in favor of an important role of this mechanism in collective effects during smooth plastic flow were advanced in [12]. However, further studies are needed to elucidate whether the same mechanism of spatial coupling by elastic stresses can control the collective dislocation dynamics in a scale range spreading from the motion of individual dislocations to the formation of large deformation bands.

Another challenge concerns the choice between the dynamical mechanisms put forward to explain the diversity of the statistics of stress serrations. The framework presented in this Chapter interprets it in terms of SOC and synchronization phenomena. According to the power-law statistics for the amplitudes and durations of AE events and Poisson statistics of interevent intervals, SOC is manifested on mesoscopic scales at all strain rates. This mechanism also controls scale-free behavior of type $A$ stress serrations at high $\dot{\varepsilon}_a$. The synchronization of dislocation avalanches implies the occurrence of characteristic scales of stress serrations when $\dot{\varepsilon}_a$ is decreased. An alternative interpretation of scale-free statistics of type $A$ serrations was proposed on the basis of a model of the PLC effect, which considers a coupled evolution of several dislocation subsystems, including mobile and forest dislocations, but also mobile dislocations dragging solute atoms [78,79]. The analysis of the Lyapunov exponents [26], which characterize the convergence (or divergence) of close phase trajectories of a dynamical system, revealed behavior similar to turbulent flow [155,156]. This model is able to predict, by implementing a unique framework, the transitions from stress serrations with a characteristic scale to scale-invariant behavior. However, this model does not seem to predict a distinction between small and large scales at low strain rates. In view of these alternative hypotheses, the definite answer as to the mechanism governing scale-free behavior related to the PLC effect needs further investigation.

An approach which is new and old at the same time has been proposed recently [157,158]. This method recalls that complex systems of various nature display a generic feature known as fluctuation scaling or Taylor's law (after investigations in ecology [159]), which relates the average and the variance of fluctuations in complex systems by a power-law. It has been argued that although numerous system-specific dynamical models were proposed with more or less success to explain the emergence of power-laws in various fields of research, SOC and (multi)fractal behavior naturally derive from this general phenomenon and are related to the convergence of a wide range of statistical processes to the so-called Tweedie distributions [160]. The first attempt to verify the fluctuation scaling in the case of the PLC effect was reported in a very recent paper which examined the statistics of type *A* stress serrations and of the accompanying local strain-rate bursts recorded by an optical technique at a frequency of 1,000 Hz [53]. This method occupies an intermediate place between the measurements of the AE and deformation curves, with regard to the sensitivity and temporal resolution. Without going into detail, it should be mentioned that the statistical distributions of the local strain rate demonstrated power-law dependences, in agreement with the scaling behaviors in the outermost scale ranges corresponding to AE and deformation curves. As far as the discussed concept of fluctuation scaling is concerned, the results of analysis showed that both time series, $\sigma(t)$ and $\dot{\varepsilon}_{loc}(t)$, obey a similar power-law with the exponent value typical of sandpile models considered as a paradigm of SOC [157,158]. More generally, this scaling behavior conforms to a certain class of complex dynamical systems characterized by compound Poisson–gamma distributions. As a matter of analogy, such statistics are particularly used to mimic the process of capturing clusters in ecological data such as biomasses [161].

### 4.2. Wave-intermittence duality on small scales

Besides completing investigations regarding the intermittency of plastic flow in an intermediate scale range, the use of a local extensometry technique in [53] has highlighted a qualitatively different aspect of self-organization of plastic flow. To present this novel facet, it should first be recalled that, as illustrated in Section *3.3* using the example of the statistical analysis of intermittency, the apparent behavior may depend on the observed quantity and on the scale of observation. The application of optical methods to investigate the local strain field on the specimen surface, e.g., digital image correlation [162] or speckle interferometry [163], has given yet a wider meaning to this statement. Such experiments revealed waves of strain localization, another ubiquitous feature of plastic flow which neither requires specific mechanisms of macroscopic instability [13,164-167]. In comparison with the intermittency revealed by the AE technique, this phenomenon corresponds to a distinct frequency region implied by the typical sampling rate of 10 frames per second. The observed waves have a wavelength about 1 to 10 mm and a low propagation velocity, usually in a range of $10^{-2}$-$10^{-1}$ mm/s, which correspond to characteristic frequencies below 0.1 Hz. It can be assumed that each of the two aspects of spatiotemporal behavior may or may not manifest, depending on the experimental technique used. The statistical analysis of the EA is usually carried out for global time series and does not consider the spatial structuring of plastic deformation. At the same time, local strain measurements most often have a fairly coarse temporal resolution and neglect the intermittent nature of the propagation of deformation. The current situation in this field of investigation is that two groups of studies are mostly isolated from each other, each giving priority to one or another aspect.

A few recent works reported on a duality between these two behaviors. Figure 34 presents a local strain-rate map, similar to Figures 4 and 5, for an early stage of plastic deformation of a Cu single crystal [12]. Despite a perfectly smooth character of its deformation curve, the $\dot{\varepsilon}(x,t)$ diagram obtained using a high-frequency (1,000 Hz) local extensometry reveals both the intermittence and the propagation of plastic flow. The intermittence is manifested by bright spots reflecting $\dot{\varepsilon}_{loc}$ bursts, as confirmed by statistical analysis that revealed a power-law character of the distributions of the burst amplitudes. At the same time, such bursts are arranged along oblique straight lines that reflect the propagation of local strain heterogeneities.

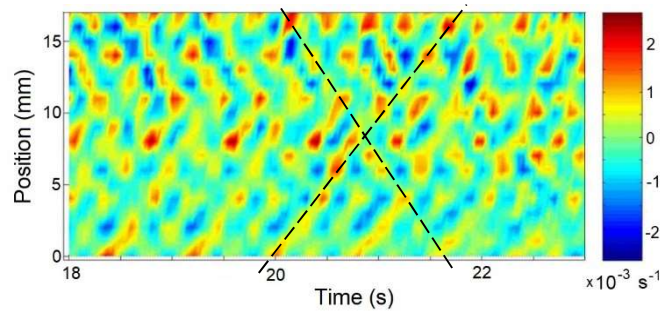

**Figure 34**. Spatiotemporal map illustrating longitudinal fluctuations about the imposed strain rate ($\dot{\varepsilon}_a = 5 \times 10^{-4} s^{-1}$) during the elastoplastic transition. Similar to Figure 4, the color bar represents the local $\dot{\varepsilon}$ scale. Fluctuations can be as high as $2.5 \times 10^{-3} s^{-1}$. Dotted characteristic lines run from the left and right of the gauge length, reflecting intermittency and propagation of strain localization (Figure adapted from Ref. [12]).

The coexistence of two dynamical modes was interpreted in the framework of a dislocation field theory which considered both the transport of dislocations involving short-range interactions of dislocations with obstacles and the spatial coupling between dislocations due to their internal stress field (Figure 35).

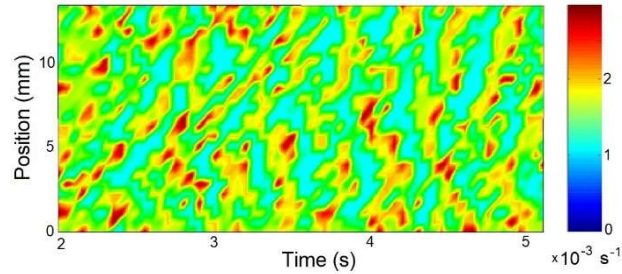

***Figure 35**. Model predictions of axial strain-rate fluctuations. The sample is a 13 × 13 mm² square in a glide plane subjected to equal shear rates of 5 × 10⁻⁴ s⁻¹ on both sides.*

Such maps were later observed in several materials with different crystal structure and defect microstructure, e.g., in α-titanium [117,168] and TWIP steel [169]. It is of interest in the context of the Chapter that the corresponding scale presents similar patterns in the conditions of the PLC effect. A hint to the presence of such patterns during jerky flow was provided by Figure 5 where weak $\dot{\varepsilon}_{loc}$ heterogeneities could be discerned in the regions between the PLC bands dominating the contrast of the colored map. A clearer view is given by Figure 36, which displays portions of a strain-rate map during "quiescent" intervals, i.e., before $\varepsilon_{cr}$ or during reloading after a stress drop [45,124].

It can be conjectured from the examples of Figures 34-36 that the coexistence of the intermittence and waves is a common property of various materials on a certain scale of deformation processes, which occurs during both smooth and jerky plastic flow. At the same time, the compliance between the two aspects evolves over the course of deformation and may alternate and even be substituted by disordered patterns (see Figure 36). The similarity between Figures 34 and 36 advances a hypothesis of a general mechanism determining self-organization of deformation processes in various materials on mesoscopic scales, in consistence with the common power-law character of statistical distributions of acoustic emission which reflects yet finer scales of plastic flow. Although the study of this mechanism is at an initial stage, it can be conjectured that the relevant behaviors are of a purely dynamical nature. The wave-intermittence duality, very little explored so far, presents a great interest for the understanding of correlations between temporal instabilities and spatial heterogeneities in the system of crystal defects. On the other hand, macroscopic instabilities, such as the PLC effect, are controlled by specific mechanisms and manifest diverse patterns. These mechanisms may have an influence on the behavior at small scales. For example, it was discussed in Section *3.3* that the power-law statistical distributions of AE are characterized by higher exponents in the case of dynamically strain ageing alloys. Developing a model combining the DSA mechanism with that of the dislocation transport and understanding the interaction between two mechanisms represents a challenge for future research in this area.

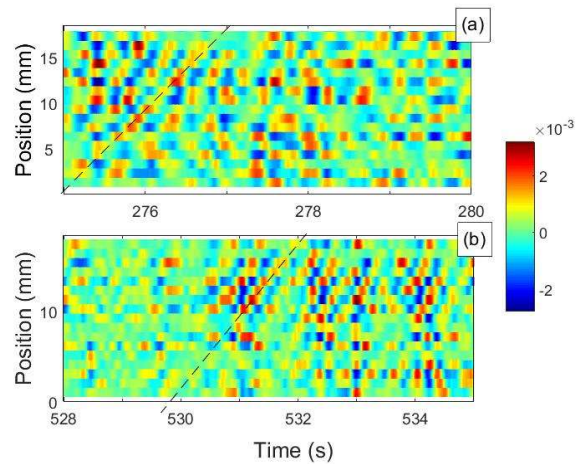

***Figure 36****. Examples of low-amplitude fluctuations of local strain rates for an AlMg sample. (a) Time interval before $\varepsilon_{cr}$; (b) Between stress serrations. $\dot{\varepsilon}_a = 2 \times 10^{-4}\,\mathrm{s}^{-1}$.*

In summary, the present Chapter considered several approaches to experimental investigation and quantitative analysis of complex behaviors emerging on different scales during the PLC effect in conventional alloys. Some aspects of the observed patterns are also common for the macroscopically smooth deformation of various crystal materials. Moreover, the approaches developed for these investigations have a general character and may be useful for investigation of a multiscale complexity of plasticity in diverse novel materials, be it related to either conventional crystal defects or to specific mechanisms of deformation. It is useful to recall in this context small-scale behaviors that were discussed in this Chapter. While macroscopic instabilities in such materials as high-entropy alloys or metallic glasses have already received much attention of researches, as presented in this book, fine behaviors that can be revealed by the AE or local strain field measurements remain largely unknown. In our opinion, this is an indispensable step to the understanding of the properties of novel materials.

## References

1. F. Savart, Recherches sur les vibrations longitudinales, Ann. Chim. Fhys. 65 (1837) 337-402.

2. M.A. Masson, Sur l'élasticité des corpes solides, Ann. Chim. Phys. 3 (1841) 451-462.

3. A. Portevin, F. Le Chatelier, Sur un phénomène observé lors de l'essai de traction d'alliages en cours de transformation, C.R. Acad. Sci. Paris. 176 (1923) 507-510.

4. Y. Estrin, L.P. Kubin, Spatial coupling and propagative plastic instabilities. In: H.-B. Mühlhaus (Ed.) Continuum models for materials with microstructure. Wiley & Sons. New York (1995) pp. 395-450.

5. G. Nicolis, I. Prigogine, Self-Organization in Nonequilibrium System: From Dissipative Structures to Order Through Fluctuations. J. Wiley & Sons. New York, London, Sydney etc., 1977. ISBN: 0-471-02401-5.

6. H. Haken, Synergetik - Eine Einführung. Springer-Verlag. Berlin, Heidelberg, New York, 1982.

7. L.P. Kubin, C. Fressengeas, G. Ananthakrishna, Collective behavior of dislocations in plasticity. In: F.R.N. Nabarro and M.S. Duesbery (Eds.) Dislocations in Solids. Elsevier Science BV. Amsterdam, Vol. 11 (2002) pp. 101-192.

8. G. Ananthakrishna, Current theoretical approaches to collective behavior of dislocations, Phys. Rep.-Rev. Sec. Phys. Lett. 440 (2007) 113-259. https://doi.org/10.1016/j.physrep.2006.10.003

9. L.P. Kubin, Dislocation patterning. In: H. Müghrabi (Ed.) Treatise in Materials Science and Technology. VCH-Weinberg, FRG (1993) pp. 137-189.

10. G.A. Malygin, Dislocation self-organization processes and crystal plasticity, Phys.-Usp. 42 (1999) 887-916. https://doi.org/10.1070/PU1999v042n09ABEH000563

11. J. Weiss, T. Richeton, F. Louchet, F. Chmelík, P. Dobron, D. Entemeyer, M. Lebyodkin, T. Lebedkina, C. Fressengeas, R.J. McDonald, Evidence for universal intermittent crystal plasticity from acoustic emission and high-resolution extensometry experiments, Phys. Rev. B 76 (2007) 224110. https://doi.org/10.1103/PhysRevB.76.224110

12. C. Fressengeas, A.J. Beaudoin, D. Entemeyer, T. Lebedkina, M. Lebyodkin, V. Taupin, Dislocation transport and intermittency in the plasticity of crystalline solids, Phys. Rev. B 79 (2009) 014108. https://doi.org/10.1103/PhysRevB.79.014108

13. L.B. Zuev, On the way of plastic flow localization in pure metals and alloys, Ann. Phys. 16 (4) (2007) 286-310. https://doi.org/10.1002/andp.200610233

14. E.N. da C. Andrade, On the viscous flow in metals and allied phenomena, Proc. Roy. Soc. A 84 (1910) 1–12.

15. D.M. Dimiduk, C. Woodward, R. LeSar, M.D. Uchic, Scale-free intermittent flow in crystal plasticity, Science 312 (2006) 1188-1190. https://doi.org/10.1126/science.1123889

16. M.V. Klassen-Nekludova, Mechanical twinning of crystals. Consultants Bureau. New York, 1964.

17. O. V. Klyavin, Physics of Plasticity of Crystals at Helium Temperatures [in Russian]. Nauka. Moscow, 1987.

18. Y. Estrin, L.P. Kubin, Plastic instabilities: phenomenology and theory, Mater. Sci. Eng. A 137 (1991) 125-134. https://doi.org/10.1016/0921-5093(91)90326-I

19. M. Zaiser, P. Hähner, Oscillatory modes of plastic deformation: Theoretical concepts, Phys. Status Solidi (b) 199 (1997) 267-330. https://doi.org/10.1002/1521-3951(199702)199:2<267::AID-PSSB267>3.0.CO;2-Q

20. B. Obst, Basic aspects of tensile properties. In: Handbook of Applied Superconductivity, B. Seeber (Ed.) IOP Publishing. Bristol and Philadelphia (1998) pp. 969-993.

21. V.V. Pustovalov, Serrated deformation of metals and alloys at low temperatures (Review), Low Temp. Phys. 34 (2008) 683-723. https://doi.org/10.1063/1.2973710

22. P. Rodriguez, Serrated plastic flow, Bull. Mater. Sci. 6 (1984) 653-663.

23. J.M. Robinson, M.P. Shaw, Microstructural and mechanical influences on dynamic strain ageing phenomena, Intern. Mater. Reviews 39 (3) (1994) 113-122. https://doi.org/10.1179/imr.1994.39.3.113

24. M.A. Lebyodkin, N.P. Kobelev, Y. Bougherira, D. Entemeyer, C. Fressengeas, V.S. Gornakov, T.A. Lebedkina, I.V. Shashkov, On the similarity of plastic flow processes during smooth and jerky flow: Statistical analysis, Acta Mater. 60 (2012) 3729–3740. https://doi.org/10.1016/j.actamat.2012.03.026

25. K. Chihab, Y. Estrin, L.P. Kubin, J. Vergnol, The kinetics of the Portevin-Le Chatelier bands in an Al-5 at%Mg alloy, Scr. Metall. 21 (1987) 203-208. https://doi.org/10.1016/0036-9748(87)90435-2

26. H.D.I. Abarbanel, R. Brown, J.J. Sidorowich, L.Sh. Tsimring, The analysis of observed chaotic data in physical systems, Rev. Mod. Phys. 65(1993) 1331-1392. https://doi.org/10.1103/RevModPhys.65.1331

27. P. Bak, C. Tang, K. Wiesenfeld, Self-organized criticality, Phys. Rev. A 38 (1988) 364-374. https://doi.org/10.1103/PhysRevA.38.364

28. S.H. Strogatz, From Kuramoto to Crawford: Exploring the onset of synchronization in populations of coupled oscillators. Phys. D 143 (2000) 1–20. https://doi.org/10.1016/S0167-2789(00)00094-4

29. A.W. Sleeswijk, Dynamic strain ageing of alloys, Acta Metall. 6 (1958) 548.

30. P.G. McCormick, Serrated yielding and the occurrence of strain gradients in an Al-Cu alloy, Scr. Metall. 6 (1972) 165–170. https://doi.org/10.1016/0036-9748(72)90218-9

31. G.A. Malygin, Dynamic interaction of impurity atmospheres with moving dislocations during serrated flow, Phys. Stat. Sol. (a) 15 (1973) 51-60. https://doi.org/10.1002/pssa.2210150104

32. A. Van den Beukel, Theory of the effect of dynamic strain aging on mechanical properties, Phys. Stat. Sol. A 30 (1975) 197-206. https://doi.org/10.1002/pssa.2210300120

33. A. Korbel, J.D. Embury, M. Hatherly, P.L. Martin, H.W. Ersbloh, Microstructural aspects of strain localization in Al-Mg alloys, Acta Metall. 34 (10) (1986) 1999-2009. https://doi.org/10.1016/0001-6160(86)90259-2

34. P. Penning, Mathematics of the Portevin-Le Chatelier effect, Acta Metall. 20 (1972) 1169-1175. https://doi.org/10.1016/0001-6160(72)90165-4

35. Y. Estrin, L.P. Kubin, Collective dislocation behavior in dilute alloys and the Portevin-Le Chatelier effect, J. Mech. Behav. Mater. 2 (1990) 255-292.

36. Rizzi E., Hähner P. On the Portevin–Le Chatelier effect: theoretical modeling and numerical results, J. Int. Plast. 20 (2004) 121–165. https://doi.org/10.1016/S0749-6419(03)00035-4

37. S.Dj. Mesarovic, Dynamic strain aging and plastic instabilities, J. Mech. Phys. Solids 43 (1995) 671-700. https://doi.org/10.1016/0022-5096(95)00010-G

38. S. Tamimi, A. Andrade-Campos, J. Pinho-da-Cruz, Modelling of the Portevin-Le Chatelier effects in aluminium alloys: a review, J. Mech. Behav. Mater. 24 (2015) 67-78.

39. P.N. Butcher, Gunn effect, Rep. Prog. Phys. 30 (1967) 97-148. https://doi.org/10.1088/0034-4885/30/1/303

40. L. J. Cuddy, W. C. Leslie, Some aspects of serrated yielding in substitutional solid solutions of iron, Acta metall. 40 (1972) 1157-1167. https://doi.org/10.1016/0001-6160(72)90164-2

41. H. Jiang, Q. Zhang, X. Chen, Z. Chen, Z. Jiang, X. Wu, J. Fan, Three types of Portevin-Le Chatelier effects: experiment and modeling, Acta Mater. 55 (2007) 2219-2228. https://doi.org/10.1016/j.actamat.2006.10.029

42. H. Ait-Amokhtar, P. Vacher, S. Boudrahem, Kinematics fields and spatial activity of Portevine-Le Chatelier bands using the digital image correlation method, Acta Mater. 54 (2006) 4365-4371. https://doi.org/10.1016/j.actamat.2006.05.028

43. A. Yilmaz, The Portevin-Le Chatelier effect: a review of experimental findings, Sci. Technol. Adv. Mater. 12 (2011) 063001. https://doi.org/10.1088/1468-6996/12/6/063001

44. D.A. Zhemchuzhnikova, Influence of the extreme grain size reduction on plastic deformation instability in an AlMg and AlMgScZr alloys, PhD Thesis, Université de Lorraine, Metz, France, 2018. https://hal.univ-lorraine.fr/tel-02141620/document

45. Y. Bougherira, D. Entemeyer, C. Fressengeas, T. Lebedkina, N. Kobelev, M. Lebyodkin, Caractérisation multi-échelle de l'hétérogénéité spatio-temporelle de l'effet Portevin-Le Chatelier, Le XIXème Congrès Français de Mécanique, CFM 09, Marseille, France, August 24-28, 2009. http://documents.ire-vues.inist.fr/bitstream/handle/2042/37048/633.pdf?sequence=1

46. M.A. Lebyodkin, T.A. Lebedkina, Multifractality and randomness in the unstable plastic flow near the lower strain-rate boundary of instability, Phys. Rev. E 77 (2008) 026111. https://doi.org/10.1103/PhysRevE.77.026111

47. R.S. Raghavan, G. Ananthakrishna, Comment on ''Scaling Behavior of the Portevin–Le Chatelier Effect in an Al-2.5%Mg Alloy'', Phys. Rev. Lett. 97 (2006) 079601. https://doi.org/10.1103/PhysRevLett.97.079601

48. M.A. Lebyodkin, T.A. Lebedkina, A. Jacques, Multifractal analysis of unstable plastic flow. Nova Science Publishers, Inc. New York, 2009.

49. M. Lebyodkin, L. Dunin-Barkowskii, Y. Brechet, Y. Estrin, L.P. Kubin, Spatio-temporal dynamics of the Portevin-Le Chatelier effect: Experiment and modelling, Acta Mater. 48 (2000) 2529-2541. https://doi.org/10.1016/S1359-6454(00)00067-7

50. M.S. Bharathi, M. Lebyodkin, G. Ananthakrishna, C. Fressengeas, L.P. Kubin, Multifractal Burst in the Spatiotemporal Dynamics of Jerky Flow, Phys. Rev. Lett. 87(16) (2001) 165508. https://doi.org/10.1103/PhysRevLett.87.165508

51. M.S. Bharathi, M. Lebyodkin, G. Ananthakrishna, C. Fressengeas, L.P. Kubin, The hidden order behind jerky flow, Acta Mater. 50 (2002) 2813-2824. https://doi.org/10.1016/S1359-6454(02)00099-X

52. T.A. Lebedkina, D.A. Zhemchuzhnikova, M.A. Lebyodkin, Correlation versus randomization of jerky flow in an AlMgScZr alloy using acoustic emission, Phys. Rev. E 97 (2018) 013001. https://doi.org/10.1103/PhysRevE.97.013001

53. M.A. Lebyodkin, Y. Bougherira, T.A. Lebedkina, D. Entemeyer, Scaling in the Local Strain-Rate Field during Jerky Flow in an Al-3%Mg Alloy, Metals 10 (2020) 134. https://doi.org/doi:10.3390/met10010134

54. M.A. Lebyodkin, Y. Brechet, Y. Estrin, L.P. Kubin, Statistics of the catastrophic slip events in the Portevin-Le Chatelier effect, Phys. Rev. Lett. 74 (1995) 4758-4761. https://doi.org/10.1103/PhysRevLett.74.4758

55. M.A. Lebyodkin, Y. Brechet, Y. Estrin, L.P. Kubin, Statistical Behaviour And Strain Localization Patterns in the Portevin-Le Chatelier Effect, Acta mater. 44 (1996) 4531-4541. https://doi.org/10.1016/1359-6454(96)00076-6

56. M. A. Lebyodkin, L.R. Dunin-Barkowskii, Critical behavior and mechanism of strain correlations under conditions of unstable plastic flow, J. Exp. Theor. Phys. 86 (1998) 993-1000. https://doi.org/10.1134/1.558571

57. M. A. Lebyodkin, L.R. Dunin-Barkowskii, Dynamic mechanism of the temperature dependence of the Portevin–Le Chatelier effect, Phys. Solid State 40 (1998) 447-452. https://doi.org/10.1134/1.1130341

58. M.A. Lebyodkin, L.R. Dunin-Barkowskii, T.A. Lebedkina, Statistical and multifractal analysis of collective dislocation processes in the Portevin-Le Chatelier effect, Phys. Mesomech. 4 (2001) 9-14.

59. R.B. Schwarz, L.L. Funk, Kinetics of the Portevin-Le Chatelier effect in Al 6061 alloy, Acta metal. 33 (1985) 295-307. https://doi.org/10.1016/0001-6160(85)90148-8

60. G.F. Xiang, Q.C. Zhang, H.W. Liu, X.P. Wu, X.Y. Ju, Time-resolved deformation measurements of the Portevin-Le Chatelier bands, Scr. Mater. 56 (2007) 721-724. https://doi.org/10.1016/j.scriptamat.2006.08.049

61. L. Casarotto, H. Dierke, R. Tutsch, H. Neuhäuser, On nucleation and propagation of PLC bands in an Al–3Mg alloy, Mater. Sci. Eng. A 527 (2009) 132-140. https://doi.org/10.1016/j.msea.2009.07.043

62. W. Tong, H. Tao, N. Zhang, L.G. Hector Jr., Time-resolved strain mapping measurements of individual Portevin–Le Chatelier deformation bands, Scripta Mater. 53 (2005) 87–92. https://doi.org/10.1016/j.scriptamat.2005.03.020

63. J. Zdunek, T. Brynk, J. Mizera, Z. Pakiela, K. J. Kurzydlowski, Digital image correlation investigation of Portevin-Le Chatelier effect in an aluminium alloy, Material characterization 59 (2008) 1429-1433. https://doi.org/10.1016/j.matchar.2008.01.004

64. M.A. Lebyodkin, D.A. Zhemchuzhnikova, T.A. Lebedkina, E.C. Aifantis, Kinematics of formation and cessation of type B deformation bands during the Portevin-Le Chatelier effect in an AlMg alloy, Results in Physics 12 (2019) 867-869. https://doi.org/10.1016/j.rinp.2018.12.067

65. A. Clauset, C. Shalizi, M. Newman, Power-law distributions in empirical data, SIAM Rev. 51 (2009) 661-703. https://doi.org/10.1137/070710111

66. A. Deluca, Á. Corral, Fitting and goodness-of-fit test of non-truncated and truncated power-law distributions, Acta Geophys. 61 (2013) 1351-1394. https://doi.org/10.2478/s11600-013-0154-9

67. W.H. Press, Flicker noises in astronomy and elsewhere, Comments on Modern Physics, Part C - Comments on Astrophysics 7 (1978) 103-119.

68. P. Bak, Chao Tang, K. Wiesenfeld, Are Earthquakes, Fractals, and $1/f$ Noise Self-organized Critical Phenomena? In: H. Takayama (Ed.) Cooperative Dynamics in Complex Physical Systems. Springer Series in Synergetics. Springer-Verlag. Berlin, Heidelberg, Vol. 43 (1988) pp. 274-279. https://doi.org/10.1007/978-3-642-74554-6_70

69. P. Bak, K. Chen, Self-organized criticality, Sci. Am. 264 (1991) 46-53. https://doi.org/10.1038/scientificamerican0191-46

70. H.J. Jensen, Self-organized criticality. Cambridge University Press. Cambridge, 1998.

71. M.B. Weissman, 1/f noise and other slow, nonexponential kinetics in condensed matter, Rev. Mod. Phys. 60 (1988) 537-571. https://doi.org/10.1103/RevModPhys.60.537

72. E. Altshuler, T. H. Johansen, Colloquium: Experiments in vortex avalanches, Rev. Mod. Phys. 76 (2004) 471-487 https://doi.org/10.1103/RevModPhys.76.471

73. N.W. Watkins, G. Pruessner, S.C. Chapman, N.B. Crosby, H.J. Jensen, 25 Years of Self-organized Criticality: Concepts and Controversies. Space Sci. Rev. 198 (2016) 3-44. https://doi.org/10.1007/s11214-015-0155-x

74. M. Zaiser, Scale invariance in plastic flow of crystalline solids, Adv. Phys. 55 (2006) 185-245. https://doi.org/10.1080/00018730600583514

75. G. Sparks, R. Maaβ, Nontrivial scaling exponents of dislocation avalanches in microplasticity, Phys. Rev. Mater. 2 (2018) 120601. https://doi.org/10.1103/PhysRevMaterials.2.120601

76. N.Q. Chinh, T. Györi, J. Gubicza, J. Lendvai, T.G. Langdon, Possible self-organized criticality in the Portevin-Le Chatelier effect during decomposition of solid solution alloys, MRS Commun. 2 (2011) 1-4. https://doi.org/10.1557/mrc.2011.25

77. J. Kertész, L.B. Kiss, The noise spectrum in the model of self-organised criticality, J. Phys. A: Math. Gen. 23 (1990) L433-L440. https://doi.org/10.1088/0305-4470/23/9/006

78. Bharathi, M.S.; Ananthakrishna, G. Chaotic and power law states in the Portevin-Le Chatelier effect, Europhys. Lett. 60 (2002) 234–240. https://doi.org/10.1209/epl/i2002-00391-2

79. Ananthakrishna, G.; Bharathi, M.S. Dynamical approach to the spatiotemporal aspects of the Portevin–Le Chatelier effect: Chaos, turbulence, and band propagation, Phys. Rev. E 70 (2004) 026111. https://doi.org/10.1103/PhysRevE.70.026111

80. J. Feder, Fractals. Plenum Press. New York, London, 1988.

81. J.-F. Muzy, E. Bacry, A. Arneodo, The multifractal formalism revisited with wavelets, Int. J. Bif. Chaos 4 (1994) 245-302. https://doi.org/10.1142/S0218127494000204

82. D. Harte, Multifractals. Chapman & Hall. London, 2001.

83. A. Arneodo, B. Audit, N. Decoster, J.-F. Muzy, C. Vaillant, Wavelet-based multifractal formalism: applications to DNA sequences, satellite images of the cloud structure and stock market data. In: A. Bunde, J. Kropp, H. J. Schellnhuber (Eds.) The Science of Disasters. Springer (2002) pp. 27-102.

84. B.B. Mandelbrot, The fractal geometry of nature. Macmillan. NY, 1983.

85. B.B. Mandelbrot, Fractals and Chaos: The Mandelbrot Set and Beyond. Springer-Verlag. NY, 2004. https://doi.org/10.1007/978-1-4757-4017-2

86. T.C. Halsey, M.H. Jensen, L.P. Kadanoff, I.Procaccia, B.I. Shraiman, Fractal measures and their singularities: The characterization of strange sets, Phys. Rev. A 33 (1986) 1141-1151. https://doi.org/10.1103/PhysRevA.33.1141

87. A.B. Chhabra, R.V. Jensen, Direct determination of the f(α) singularity spectrum, Phys. Rev. Lett. 62 (1989) 1327-1330. https://doi.org/10.1103/PhysRevLett.62.1327

88. G. Ananthakrishna, C. Fressengeas, M. Grosbras, J. Vergnol, G. Engelke, J. Plessing, H. Neuhäuser, E. Bouchaud, J. Planes, L.P. Kubin, On the existence of chaos in jerky flow of single crystals, Scr. Metall. Mater. 32 (1995) 1731-1737. https://doi.org/10.1016/0956-716X(95)00013-L

89. G. Ananthakrishna, S.J. Noronha, C. Fressengeas, L.P. Kubin, Crossover from chaotic to self-organized critical dynamics in jerky flow of single crystals, Phys. Rev. E 60 (1999) 5455-5462. https://doi.org/10.1103/PhysRevE.60.5455

90. S. J. Noronha, G. Ananthakrishna, L. Quaouire, C. Fressengeas, L.P. Kubin, Chaos in the Portevin-Le Chatelier effect, Int. J. Bifurcation Chaos Appl. Sci. Eng. 7, 2577-2586 (1997). https://doi.org/10.1142/S0218127497001734

91. D. Kugiumtzis, A. Kehagias, E.C. Aifantis, H. Neuhäuser, Statistical analysis of the extreme values of stress time series from the Portevin - Le Chatelier effect, Phys Rev E 70 (2004) 036110. https://doi.org/10.1103/PhysRevE.70.036110

92. A. Sarkar, P. Barat, The Portevin-Le Chatelier effect in the continuous time random walk framework, Phys. Lett. A 367 (2007) 291-294. https://doi.org/10.1016/j.physleta.2007.03.020

93. A. Sarkar, C.L. Webber Jr., P. Barat, P. Mukherjee, Recurrence analysis of the Portevin–Le Chatelier effect, Physics Letters A 372 (2008) 1101–1105. https://doi.org/10.1016/j.physleta.2007.08.055

94. K. Darowicki, J. Orlikowski, A. Zielinski, W. Jurczak, Quadratic Cohen representations in spectral analysis of serration process in Al-Mg alloys, Comput. Mater. Sci. 39 (2007) 880-886. https://doi.org/10.1016/j.commatsci.2006.10.011

95. K. Darowicki, J. Orlikowski, A. Zielinski, Frequency bands selection of the Portevin-LeChatelier effect, 43 (2008) 366-373. https://doi.org/10.1016/j.commatsci.2007.12.001

96. R.N. Mudrock, M.A. Lebyodkin, P. Kurath, A.J. Beaudoin, T.A. Lebedkina, Strain-rate fluctuation during macroscopically uniform deformation of a solution-strengthened alloy. Scr. Mater. 65 (2011) 1093–1096. https://doi.org/10.1016/j.scriptamat.2011.09.025

97. A.C. Iliopoulos, N.S. Nikolaidis, E.C. Aifantis, Portevin-Le Chatelier effect and Tsallis nonextensive statistics, Physica A 438 (2015) 509. https://doi.org/10.1016/j.physa.2015.07.007

98. J. Brechtl, S.Y. Chen, X. Xie, Y. Ren, J.W. Qiao, P.K. Liaw, S.J. Zinkle, Towards a greater understanding of serrated flow in an Al-containing high-entropy-based alloy. Int. J. Plast. 115 (2019) 71–92. https://doi.org/10.1016/j.ijplas.2018.11.011

99. J. Brechtl, B. Chen, X. Xie, Y. Ren, J.D. Venable, P.K. Liaw, S.J. Zinkle, Entropy modeling on serrated flows in carburized steels. Mater. Sci. Eng. A 753 (2019) 135–145. https://doi.org/10.1016/j.msea.2019.02.096

100. E.N. Lorenz, Deterministic nonperiodic flow, J. Atmos. Sci. 20 (1963) 130-141. https://doi.org/10.1175/1520-0469(1963)020<0130:DNF>2.0.CO;2

101. G. Ananthakrishna, M.C. Valsakumar, Chaotic flow in a model for repeated yielding, Phys. Lett. A 95 (1983) 69-71. https://doi.org/10.1016/0375-9601(83)90141-X

102. S. Kok, M.S. Bharathi, A.J. Beaudoin, C. Fressengeas, G. Ananthakrishna, L.P. Kubin, M.A. Lebyodkin, Acta Mater. 51 (2003) 3651-3662. https://doi.org/10.1016/S1359-6454(03)00114-9

103. R. Sarmah, G. Ananthakrishna, Influence of system size on spatiotemporal dynamics of a model for plastic instability: Projecting low-dimensional and extensive chaos, Phys. Rev. E 87 (2013) 052907. https://doi.org/10.1103/PhysRevE.87.052907

104. M.A. Lebyodkin, T.A. Lebedkina, Multifractal analysis of evolving noise associated with unstable plastic flow, Phys. Rev. E 73 (2006) 036114. https://doi.org/10.1103/PhysRevE.73.036114

105. M.A. Lebyodkin, Y. Estrin, Multifractal analysis of the Portevin-Le Chatelier effect: General approach and application to AlMg and AlMg/Al2O3 alloys, Acta Mater. 53 (2005) 3403-3413. https://doi.org/10.1016/j.actamat.2005.03.042

106. A.A. Shibkov, M.F. Gasanov, M.A. Zheltov, A.E. Zolotov, V.I. Ivolgin, Intermittent plasticity associated with the spatio-temporal dynamics of deformation bands during creep tests in an AlMg polycrystal, Int. J. Plast. 86 (2016) 37-55. https://doi.org/10.1016/j.ijplas.2016.07.014.

107. F.B. Klose, F. Hagemann, P. Hähner, H. Neuhäuser, Investigation of the Portevin-Le Chatelier effect in Al-3wt.%Mg alloys by strain-rate and stress-rate controlled tensile tests, Mater. Sci. Eng. A 387–389 (2004) 93–97. https://doi.org/10.1016/j.msea.2004.01.062

108. M.M. Krishtal, A.K. Khrustalev, A.V. Volkov, S.A. Borodin, Nucleation and growth of macrofluctuations of plastic strain with discontinuous yield and luders deformation: Results of high-speed video filming, Dokl. Phys. 54 (2009) 225-229. https://doi.org/10.1134/S1028335809050024

109. A.A. Shibkov, M.A. Zheltov, M.F. Gasanov, A.E. Zolotov, A.A. Denisov, M.A. Lebyodkin, Mater. Sci. Eng. A. 2020. V. 772. P. 138777. https://doi.org/10.1016/j.msea.2019.138777

110. T. Böhlke, G. Bondar, Y. Estrin, M. Lebyodkin, Geometrically non-linear modeling of the Portevin-Le Chatelier effect, Comput. Mater. Sci. 44 (2009) 1076-1088. https://doi.org/10.1016/j.commatsci.2008.07.036

111. M. Maziere, H. Dierke, Investigations on the Portevin-Le Chatelier critical strain in an aluminum alloy, Comput. Mater. Sci. 52 (2012) 68-72. https://doi.org/10.1016/j.commatsci.2011.05.039

112. Acoustic emission testing: Basics for research – applications in civil engineering, Ch.U. Grosse, M. Ohtsu (Eds.) Springer-Verlag. Berlin, Heidelberg, 2008. https://doi.org/10.1007/978-3-540-69972-9

113. K. Ono, Acoustic emission in materials research – a review, J. Acoustic Emission, 29 (2011) 284-308.

114. Acoustic emission, W. Sikorski (Ed.) InTech. Rijeka, Croatia, 2012.

115. P. Lukac, Z. Trojanova, F. Chmelik, Microstructural characterization by non-destructive methods, Mater. Sci. Forum 482 (2005) 103-108. https://doi.org/10.4028/www.scientific.net/MSF.482.103

116. F. Chmelik, F.B. Klose, H. Dierke, J. Sachl, H. Neuhäuser, P. Lucas, Investigating the Portevin-Le Chatelier effect in strain rate and stress rate controlled tests by the acoustic emission and laser extensometry techniques, Mater. Sci. Eng. A 462 (2007) 53-60. https://doi.org/10.1016/j.msea.2006.01.169

117. M. Lebyodkin, K. Amouzou, T. Lebedkina, T. Richeton, A. Roth, Complexity and anisotropy of plastic flow of $\alpha$-Ti probed by acoustic emission and local extensometry. Materials 11 (2018) 1061. https://doi.org/10.3390/ma11071061

118. M.-C. Miguel, A. Vespignani, S. Zapperi, J. Weiss, J.-R. Grasso, Intermittent dislocation flow in viscoplastic deformation, Nature 410 (2001) 667-671. https://doi.org/10.1038/35070524

119. J. Weiss, J.-R. Grasso, Acoustic emission in single crystals of ice, J. Phys. Chem. B. 101 (1997) 6113-6117. https://doi.org/10.1021/jp963157f

120. J. Weiss, J.-R. Grasso, M.-C. Miguel, A. Vespignani, S. Zapperi, Complexity in dislocation dynamics: experiments, Mater. Sci. Eng. A. 309-310 (2001) 360-364. https://doi.org/10.1016/S0921-5093(00)01633-6

121. J. Weiss, W. Ben Rhouma, T. Richeton, S. Dechanel, F. Louchet, L. Truskinovsky, From Mild to WildLV14378 Fluctuations in Crystal Plasticity, Phys. Rev. Lett. 114 (2015) 105504. https://doi.org/10.1103/PhysRevLett.114.105504

122. V.S. Bobrov, M.A. Lebedkin, Electrical effects in low-temperature abrupt deformation of al, Fiz. Tverd. Tela 31 (1989) 120-126 (1989) [Sov. Phys. Solid State 31 (1989) 982-986]

123. V.S. Bobrov, S.I. Zaitsev, M.A. Lebedkin, Statistics of dynamic processes in the low-temperature discontinuous deformation of metals, Fiz. Tverd. Tela 32 (1990) 3060-3065. [Sov. Phys. Solid State 32 (1990) 1176-1779].

124. Y. Bougherira, Étude des phénomènes d'auto-organisation des ensembles de dislocations dans un alliage au vieillissement dynamique, PhD Thesis, Université Paul Verlaine, Metz, France, 2011. https://hal.univ-lorraine.fr/tel-01749084/document

125. I.V. Shashkov, Multiscale study of the intermittency of plastic deformation by acoustic mission method, PhD Thesis, Université de Lorraine, Metz, France, 2012. https://hal.univ-lorraine.fr/tel-01749405/document

126. K. Ono. Review on Structural Health Evaluation with Acoustic Emission, Appl. Sci. 8 (2018) 958. https://doi.org/10.3390/app8060958

127. E. Onajite, Understanding Sample Data, in: Seismic data analysis in hydrocarbon exploration. Elsevier B.V. (2014) pp. 105-115. https://doi.org/10.1016/B978-0-12-420023-4.00008-3

128. M.A. Lebyodkin, I.V. Shashkov, T.A. Lebedkina, K. Mathis, P. Dobron, F. Chmelik, Role of superposition of dislocation avalanches in the statistics of acoustic emission during plastic deformation. Phys. Rev. E 88 (2013) 042402. https://doi.org/10.1103/PhysRevE.88.042402

129. M.A. Lebyodkin, I.V. Shashkov, T.A. Lebedkina, V.S. Gornakov, Experimental investigation of the effect of thresholding on temporal statistics of avalanches, Phys. Rev. E. 95 (2017) 032910. https://doi.org/10.1103/PhysRevE.95.032910

130. I.V. Shashkov, M.A. Lebyodkin, T.A. Lebedkina, Multiscale study of acoustic emission during smooth and jerky flow in an AlMg alloy, Acta. Mater. 60 (2012) 6842-6850. https://doi.org/10.1016/j.actamat.2012.08.058

131. N. Kiesewetter, P. Schiller, Acoustic-emission from moving dislocations in aluminum, Phys. Status Solidi A 38 (1976) 569-575. https://doi.org/10.1002/pssa.2210380218

132. T.A. Lebedkina, Y. Bougherira, D. Entemeyer, M.A. Lebyodkin, I.V. Shashkov, Crossover in the scale-free statistics of acoustic emission associated with the Portevin-Le Chatelier instability, Scr. Mater. 148 (2018) 47–50. DOI: 10.1016/j.scriptamat.2018.01.017

133. M.A. Lebyodkin, T.A. Lebedkina, F. Chmelik, T.T. Lamark, Y. Estrin, C. Fressengeas, J. Weiss, Intrinsic structure of acoustic emission events during jerky flow in an Al alloy, Phys. Rev. B 79 (2009) 174114-9. https://doi.org/10.1103/PhysRevB.79.174114

134. F. Chmelík, A. Ziegenbein, H. Neuhäuser H, P. Lukáč, Investigating the Portevin-Le Chatelier effect by the acoustic emission and laser extensometry techniques, Mater. Sci. Eng. A 324 (2002) 200-207. https://doi.org/10.1016/S0921-5093(01)01312-0

135. N.P. Kobelev, M.A. Lebyodkin, T.A. Lebedkina, Role of Self-Organization of Dislocations in the Onset and Kinetics of Macroscopic Plastic Instability, Metall. Mater. Trans. A 48 (2017) 965-974. https://doi.org/10.1007/s11661-016-3912-x

136. M. Abbadi, P. Hähner, A. Zeghloul, On the characteristics of Portevin-Le Chatelier bands in aluminum alloy 5182 under stress-controlled and strain-controlled tensile testing, Mater. Sci. Eng. A 337 (2002) 194-201. https://doi.org/10.1016/S0921-5093(02)00036-9

137. K. Mathis, F. Chmelik, Exploring plastic deformation of metallic materials by the acoustic emission technique, in: W. Sikorsky (Ed.), Acoustic Emission, InTech, Rijeka 2012, pp. 23–48.

138. K. Kumar, G. Ananthakrishna, Modeling the complexity of acoustic emission during intermittent plastic deformation: Power laws and multifractal spectra, Phys. Rev. E 97 (2018) 012201. https://doi.org/10.1103/PhysRevE.97.012201

139. J.M. Carlson, J.S. Langer, B.E. Shaw, Dynamics of earthquake faults. Rev. Mod. Phys. 66 (1994) 657-670. https://doi.org/10.1103/RevModPhys.66.657

140. G. Boffetta, V. Carbone, P. Giuliani, P. Veltri, A. Vulpiani, Power laws in solar flares: Self-organized criticality or turbulence? Phys. Rev. Lett. 83 (1999) 4662-4665. https://doi.org/10.1103/PhysRevLett.83.466

141. E. Spada, V. Carbone, R. Cavazzana, L. Fattorini, G. Regnoli, N. Vianello, V. Antoni, E. Martines, G. Serianni, M. Spolaore, L. Tramontin, Search of self-organized criticality processes in magnetically confined plasmas: Hints from the reversed field pinch configuration, Phys. Rev. Lett. 86 (2001) 3032-3035. https://doi.org/10.1103/PhysRevLett.86.3032

142. L. I. Salminen, A. I. Tolvanen, M. J. Alava, Acoustic emission from paper fracture, Phys. Rev. Lett. 89 (2002) 185503. https://doi.org/10.1103/PhysRevLett.89.185503

143. A. Deluca, Á. Corral, Scale invariant events and dry spells for medium-resolution local rain data, Nonlin. Proc. Geophys. 21 (2014) 555-567. https://doi.org/10.5194/npg-21-555-2014

144. R. Sanchez, D. E. Newman, B. A. Carreras, Waiting-time statistics of self-organized-criticality systems, Phys. Rev. Lett. 88 (2002) 068302. https://doi.org/10.1103/PhysRevLett.88.068302

145. J. Baró, Á. Corral, X. Illa, A. Planes, E. K. H. Salje, W. Schranz, D. E. Soto-Parra, E. Vives, Statistical Similarity between the Compression of a Porous Material and Earthquakes, Phys. Rev. Lett. 110 (2013) 088702. https://doi.org/10.1103/PhysRevLett.110.088702

146. K. Christensen, Z. Olami, Variation of the Gutenberg-Richter b values and nontrivial temporal correlations in a spring-block model for earthquakes, J. Geophys. Res. 97 (1992) 8729-8735. https://doi.org/10.1029/92JB00427

147. M. Paczuski, S. Boettcher, M. Baiesi, Interoccurrence times in the Bak-Tang-Wiesenfeld sandpile model: A comparison with the observed statistics of solar flares, Phys. Rev. Lett. 95 (2005) 181102. https://doi.org/10.1103/PhysRevLett.95.181102

148. L. Laurson, X. Illa, M. J. Alava, The effect of thresholding on temporal avalanche statistics, J. Stat. Mech.-Theory Exp. (2009) P01019. https://doi.org/10.1088/1742-5468/2009/01/P01019

149. D. Zhemchuzhnikova, M. Lebyodkin, T. Lebedkina, A. Mogucheva, D. Yuzbekova, R. Kaibyshev, Peculiar Spatiotemporal Behavior of Unstable Plastic Flow in an AlMgMnScZr Alloy with Coarse and Ultrafine Grains, Metals 7 (2017) 325. https://doi.org/10.3390/met7090325

150. J. Buck, E. Buck, Synchronous fireflies, Sci. Am. 234 (1976) 74.

151. C.J. Pérez, Á. Corral, A. Díaz-Guilera, K. Christensen, A. Arenas, On self-organized critically and synchronization in lattice models of coupled dynamic systems, Int. J. Mod. Phys. B 10 (1996) 1111-1151. https://doi.org/10.1142/S0217979296000416

152. P.G. McCormick, C.P. Ling, Numerical modelling of the Portevin-Le Chatelier effect, Acta Metall. Mater. 43 (1995) 1969-1977. https://doi.org/10.1016/0956-7151(94)00390-4

153. M.A. Lebyodkin, N.P. Kobelev, Y. Bougherira, D. Entemeyer, C. Fressengeas, T.A. Lebedkina, I.V. Shashkov, On the similarity of plastic flow processes during smooth and jerky flow in dilute alloys, Acta Mater. 60 (2012) 844-850. https://doi.org/10.1016/j.actamat.2011.10.042

154. R. Král, P. Dobroň, F. Chmelík, V. Koula, M. Rydlo, M. Janeček, A qualitatively new approach to acoustic emission measurements and its application to pure aluminium and Mg-Al alloys, Kovove Mater. 45 (2007) 159-163.

155. F. Heslot, B. Castaing, A. Libchaber, Transitions to turbulence in helium gas, Phys. Rev, A 36 (1987) 5870-5873. https://doi.org/10.1103/PhysRevA.36.5870

156. M. Yamada, K. Ohkitani, Lyapunov spectrum of a model of two-dimensional turbulence, Phys. Rev. Lett. 60 (1988) 983-986. https://doi.org/10.1103/PhysRevLett.60.983

157. Z. Eisler, I. Bartos, J. Kertész, Fluctuation scaling in complex systems: Taylor's law and beyond, Adv. Phys. 57 (2008) 89-142. https://doi.org/10.1080/00018730801893043

158. W.S. Kendal, Self-organized criticality attributed to a central limit-like convergence effect, Physica A 421 (2015) 141-150. https://doi.org/10.1016/j.physa.2014.11.035

159. L.R. Taylor, Aggregation, variance and mean, Nature 189 (1961) 732-735. https://doi.org/10.1038/189732a0

160. B. Jørgensen, J.R. Martinez, M. Tsao, Asymptotic-behavior of the variance function, Scand. J. Stat. 21 (1994) 223-243.

161. J.-B. Lecomte, H.P. Benoît, S. Ancelet, M.-P. Etienne, L. Bel, E. Parent, Compound Poisson-gamma vs. delta-gamma to handle zero-inflated continuous data under a variable sampling volume, Methods Ecol. Evol. 4 (2013) 1159-1166. https://doi.org/10.1111/2041-210X.12122

162. M.A. Sutton, F. Hild, Recent Advances and Perspectives in Digital Image Correlation, Exp. Mech. 55 (2015), 1-8. https://doi.org/10.1007/s11340-015-9991-6

163. P. Jacquot, Speckle Interferometry: A Review of the Principal Methods in Use for Experimental Mechanics Applications, Strain 44 (2008) 57-69. https://doi.org/10.1111/j.1475-1305.2008.00372.x

164. L.B. Zuev, V.I. Danilov, N.V. Kartashova, S.A. Barannikova, The self-excited wave nature of the instability and localization of plastic deformation, Mater. Sci. Eng. A 234-236 (1997) 699-702. https://doi.org/10.1016/S0921-5093(97)00242-6

165. G.F. Sarafanov, Plastic-strain-softening waves in crystals, Phys. Solid State 43 (2001) 263-269. https://doi.org/10.1134/1.1349472

166. R.J. McDonald, C. Efstathiou, P. Kurath, The wavelike plastic deformation of single crystal copper, J. Eng. Mater. Technol. Trans. ASME 131 (2009) 031013. https://doi.org/10.1115/1.3120410

167. L.B. Zuev, S.A. Barannikova, Autowave physics of material plasticity, Crystals 9 (2019) 458. https://doi.org/10.3390/cryst9090458

168. A. Roth, M.A. Lebyodkin, T.A. Lebedkina, J.S. Lecomte, T. Richeton, K.E.K. Amouzou, Mechanisms of anisotropy of mechanical properties of alpha-titanium in tension conditions, Mater. Sci. Eng. A 596 (2014) 236-243. https://doi.org/10.1016/j.msea.2013.12.061

169. A. Roth, T.A. Lebedkina, M.A. Lebyodkin, On the critical strain for the onset of plastic instability in an austenitic FeMnC steel. Mater. Sci. Eng. A 539 (2012) 280-284. https://doi.org/10.1016/j.msea.2012.01.094